\documentclass[aps,prb,twocolumn,showpacs,superscriptaddress]{revtex4-1}
\usepackage{graphicx}
\usepackage{dcolumn}
\usepackage{amsmath}
\usepackage{amssymb}
\usepackage{color}
\usepackage{subfigure,amsmath,verbatim,moreverb,bm}
\usepackage{hyperref}
\def\be{\begin{equation}}
\def\ee{\end{equation}}
\def\ber{\begin{eqnarray}}
\def\eer{\end{eqnarray}}
\def\bmat{\begin{pmatrix}}
\def\emat{\begin{pmatrix}}

\def\bc{}

\def\xv{{\bf x}}
\def\rv{{\bf r}}

\def\jv{{\bf j}}

\def\Av{{\bf A}}
\def\Bv{{\bf B}}
\def\Ev{{\bf E}}

\def\vv{{\bf v}}
\def\Sv{{\bf S}}

\def\id{\mathrm{i}}
\def\nn{\nonumber}

\def\Im{{\rm Im}}
\def\Re{{\rm Re}}
\def \SE {Schr\"odinger equation~}
\def \SEs {Schr\"odinger equations~}

\def	 \Q {Eq.~}
\def	 \Qs {Eqs.~}
\def \d {\mathrm{d}}
\def \UP{\uparrow}
\def \DN{\downarrow}
\def \B{ \mathcal{B}}

\def \H {\mathcal{H}}

\begin{document}
\title{Dynamics of observables and exactly solvable quantum problems: Using time-dependent density functional theory to control quantum systems}
\author{M. Farzanehpour}
\email{m.farzanehpour@gmail.com}
\affiliation{Nano-Bio Spectroscopy group and ETSF Scientific Development Centre, 
  Departamento de F\'isica de Materiales, Universidad del Pa\'is Vasco UPV/EHU, E-20018 San Sebasti\'an, Spain}

\author{ I. V. Tokatly}
\email{ilya.tokatly@ehu.es}
\affiliation{Nano-Bio Spectroscopy group and ETSF Scientific Development Centre, 
  Departamento de F\'isica de Materiales, Universidad del Pa\'is Vasco UPV/EHU, E-20018 San Sebasti\'an, Spain}
\affiliation{IKERBASQUE, Basque Foundation for Science, E-48011 Bilbao, Spain}

\date{\today}

\begin{abstract}
We use analytic (current) density-potential maps of time-dependent (current) density functional theory (TD(C)DFT) to inverse engineer analytically solvable time-dependent quantum problems. In this approach the driving potential (the control signal) and the corresponding solution of the \SE are parametrized analytically in terms of the basic TD(C)DFT observables. We describe the general reconstruction strategy and illustrate it with a number of explicit examples. First we consider the real space one-particle dynamics driven by a time-dependent electromagnetic field and recover, from the general TDDFT reconstruction formulas, the known exact solution for a driven oscillator with a time-dependent frequency. Then we use analytic maps of the lattice TD(C)DFT to control quantum dynamics in a discrete space. As a first example we construct a time-dependent potential which generates prescribed dynamics on a tight-binding chain. Then our method is applied to the dynamics of spin-1/2 driven by a time dependent 
magnetic field. We design an analytic control pulse that transfers the system from the ground to excited state and vice versa. This pulse generates the spin flip thus operating as a quantum NOT gate.
\end{abstract} 

\maketitle

\section{Introduction}

Exact analytic solutions of the \SE have always been of great methodological interest as they underlie our intuitive understanding of quantum systems. Numerous solutions of the stationary \SE are known since the early days of quantum mechanics \cite{LandauIII}. However, for a long time there were very few examples of analytic solutions describing an evolution of a quantum system driven by time-dependent external field, like solutions of the Landau-Zener \cite{Landau1932,Zener1932} and Rabi  \cite{Rabi1937} problems, or the solution for a driven harmonic oscillator \cite{PopPer1969,PopPer1970} closely related to a so called harmonic potential theorem \cite{Dobson1994,Vignale1995a,Vignale1995b} . The interest in analytic solutions to time-dependent quantum problems was renewed with the emergence of quantum computing.  The necessity of designing quantum gates requires an accurate control of qubit dynamics \cite{EcoSopetal2006,GreiEcoetal2009,PoeKenetal2011} and the state preparation  \cite{WuPipetal2011,
BriCreetal2012,MalBasetal2013}. It has been recognized that by using analytical pulses in quantum control problems one achieves a more robust evolution against errors and pulse parameters \cite{EcoSopetal2006,MotGametal2009,Economou2012,MotGametal2009,ChoDicetal2010,GamMotetal2011}, which explains the practical importance of finding new solvable quantum problems and the popularity of a few known pulses for the analytic control of two-level systems. \cite{BamBer1981,BamArtetal1984,Hioe1984,Zakarzewski1985,SilJosHou1985,Robinson1985,Ishkhanyan2000,KyoVit2005,Vitanov20007}. 

In the last decade a number of new analytic solutions to the \SE have been constructed by inverse engineering time-dependent Hamiltonians from given dynamics of state vectors. We note that most of these studies focus on dynamics of two-level systems \cite{Ishkhanyan2000,GanDzeGal2010,EdwDas2012,Barnes2013,BanCheMugShe} and a few examples of three-level systems. \cite{Ishkhanyan32000} In the present work we propose an alternative strategy of reconstructing time-dependent driving potentials for analytically solvable quantum problems. Our proposal employs the ideas and theorems of time-dependent density functional theory (TDDFT) and time-dependent current density functional theory (TDCDFT) \cite{RunGro1984,TDDFT-2012,Ullrich-book}.

Originally TDDFT/TDCDFT was developed as an extension of the static DFT \cite{DreizlerGross1990} for addressing time-dependent quantum many-body problems. The key statement underlying this approach is a so called mapping theorem that establishes a one-to-one map from the time-dependent density/current to the external driving potential. The existence of this map implies that the knowledge of some properly chosen one particle observables (collective variables, such as the density or the current) is sufficient to uniquely reconstruct the corresponding conjugated driving fields, and, therefore, the full wave function of the system. For a general many-particle system the density-potential map is practically never known explicitly. In come simple situations it can be constructed numerically. For example, recently a numerical fixed-point algorithm has been used to reconstruct a potential that produces a prescribed time-dependent density in a model system.\cite{NieRugVan2013,NieRugVan2014} 

Remarkably, for one-particle systems the density-potential and the current-vector potential maps can be found explicitly in a closed analytic form \cite{MaiBurWoo2002,LiUll2008,TokatlyUni2011,TokatlyL2011,FarTok2012}. In the present work we use the explicit TDCDFT maps to construct analytic control signals driving a system in such a way that the prescribed behavior of the basic collective variable, the current and/or the density, is reproduced. The time dependence of the control signal and the dynamics of the wave function are then parametrized in terms of the physically intuitive observable. The analytic TDCDFT maps are known both for a particle in the real continuum space and for discrete, lattice (e.~g. tight binding) systems. This allows us to address, within a common scheme, control problems for the real space dynamics and for dynamics of discrete systems with a finite dimensional Hilbert space, such as a motion of quantum particle on tight-binding lattices, or the dynamics of a spin in the presence of 
a time-dependent magnetic field. To illustrate our strategy of inverse engineering we will recover the known exact solution for a driven harmonic oscillator \cite{PopPer1969,PopPer1970}, and present nontrivial examples of analytic control for a particle on a finite 1D chain and for a spin-1/2 (qubit) system.

The structure the paper is the following.  In Sec. \ref{Sec-General-idea} we present the general idea of reconstructing driving potentials for solvable problems using analytic TD(C)DFT maps. In Sec. \ref{Sec-Real-space} we use  TDCDFT to construct solvable problems for the real space one-particle dynamics. As a particular example we reconstruct the potential and the wave function generated by a density evolution in a form of a time-dependent rescaling of some initial distribution supplemented with a rigid shift in space. The corresponding solution recovers the one for the driven harmonic oscillator \cite{PopPer1969,PopPer1970}. In Sec. \ref{Sec-Discrete-space} the formalism for discrete spaces is presented. In the first subsection we give an explicit example for controlling motion of a particle on atomic chain. In the second subsection the formalism is appied to a spin-1/2 control that is isomorphic to the control problem for a particle on a two-site lattice. Finally, in Sec.\ref{Sec-Conclusion} we summarize 
our results.

\section{Construction of solvable problems via TDDFT maps: The basic idea} \label{Sec-General-idea}

\begin{figure}
  \centering
  \includegraphics[width=0.48\textwidth]{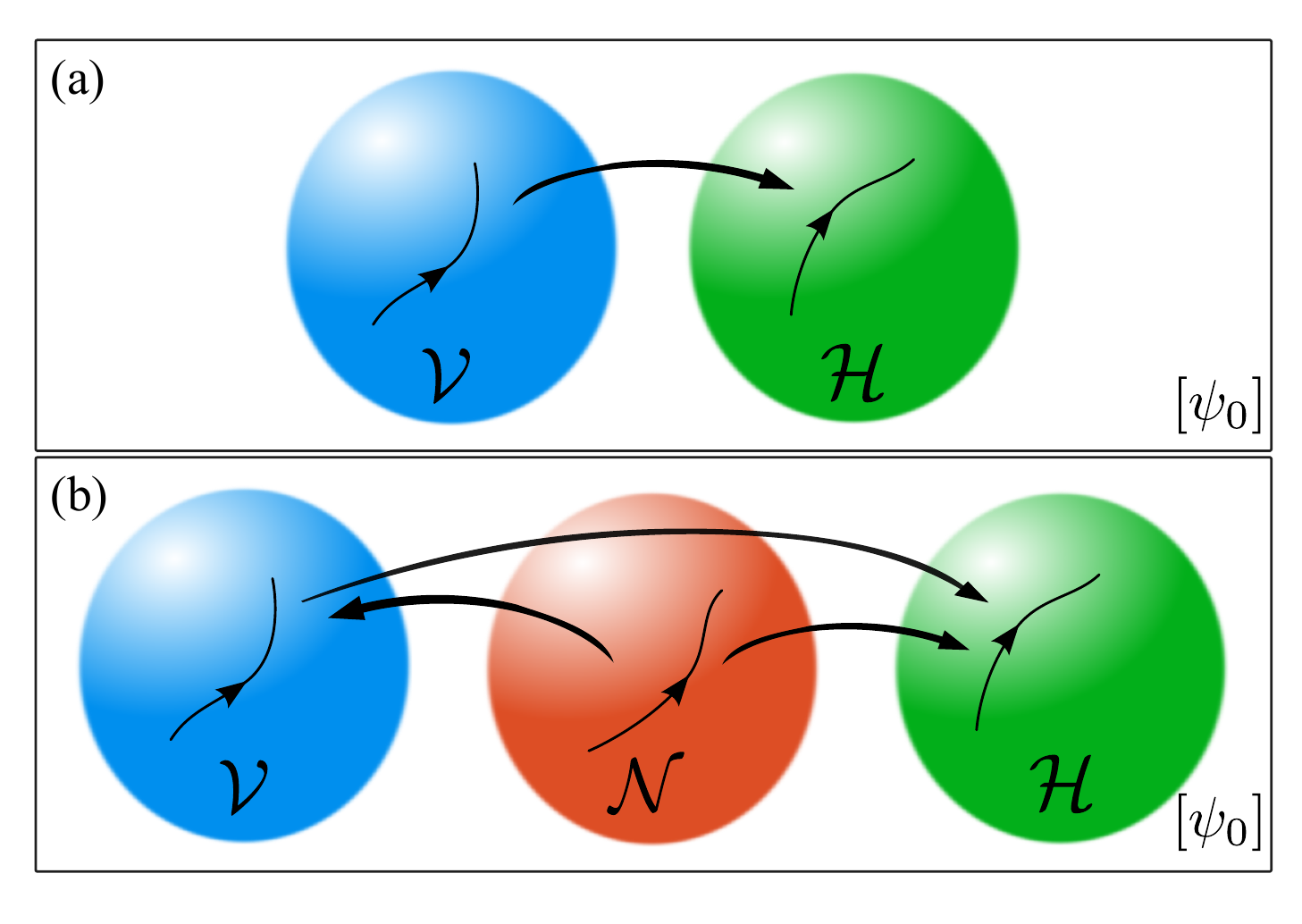}
  \caption{(a) The direct map from a trajectory in the space $\mathcal{V}$ of potentials to the trajectory in the Hilbert space $\H$, generated by the time-dependent \SE for a given initial state $\psi_0$. (b) TDDFT mappings between trajectories in the space $\mathcal{N}$ of observables (densities or currents), space $\mathcal{V}$ of potentials and the Hilbert space $\H$. For a given initial state, by choosing a desired time evolution of the density/current we can reconstruct the driving potential and the wave function. }
   \label{fig:Map}
\end{figure}

In the standard "direct" statement of a quantum mechanical problem the \SE determines evolution of the wave function $\psi(t)$ from the initial state $\psi_0$ in the presence of a given time-dependent external potential. Thus for a given initial state the \SE generates a map $\mathcal{V}\mapsto\mathcal{H}$ from the space $\mathcal{V}$ of external potentials to the Hilbert space $\mathcal{H}$. This direct map is shown schematically on Fig.~\ref{fig:Map}a. Unfortunately, the solution of the time-dependent \SE, even for simplest two-level systems, practically always requires numerical calculations. Analytically solvable time-dependent quantum problems are exceptionally rare.

To understand how TDDFT can help in finding analytically solvable problems we analyze mapping between different sets of object entering this approach. All TDDFT-type theories rely on the existence of a unique solution to a special ``inverse'' quantum problem. That is a possibility to uniquely reconstruct the driving field from a given evolution of the conjugated observable (such as, the density in TDDFT or the current in TDCDFT) and a given initial state. In other words, if $\mathcal{N}$ is the space of basic observables, then the existence of TDDFT implies that for a given initial state there exist two unique maps $\mathcal{N}\mapsto\mathcal{V}$ and $\mathcal{N}\mapsto\mathcal{H}$ which relate a given trajectory in the space $\mathcal{N}$ of observables to the corresponding trajectories in $\mathcal{V}$ and in $\mathcal{H}$. The composition of these TDDFT maps recovers the usual direct map $\mathcal{V}\mapsto\mathcal{H}$ generated by the time-dependent \SE. 

In general for many-particle systems the solution of the inverse problem is even more difficult than the solution of the usual \SE. In fact, mathematically construction of the TDDFT maps is equivalent to solving a certain nonlinear quantum many-body problem \cite{TokatlyUni2011}. However there special situations when the inverse problem possesses a simple analytic solution. These situations cover, in particular, generic driven one-particle dynamics both in the real space and on lattices with some mild restrictions on allowed initial states and the behavior of observables. For those cases the TD(C)DFT maps $\mathcal{N}\mapsto\mathcal{V}$ and $\mathcal{N}\mapsto\mathcal{H}$ can be found explicitly in the analytic form \cite{MaiBurWoo2002,LiUll2008,TokatlyUni2011,TokatlyL2011,FarTok2012}. For example, in the case of a particle driven by a time-dependent vector potential $\bf{A}(t)$ the wave function and the vector potential are the explicit analytic functionals of the current density $\jv(\rv,t)$ and the 
initial state, that is $\Av[\jv,\psi_0]$ and $\psi[\jv,\psi_0]$. By construction the wave function and the potential, obtained in such a way, are connected by the \SE with the proper initial condition. Hence by assuming different space-time distributions of the observable we can generate infinitely many solutions to the \SE, where the potential in the Hamiltonian and the solution are expressed analytically in term of the prescribed observable. In this setup the space of observables plays a role of the parameter space, while the TD(C)DFT maps provide us with the analytic parametrization formulas for the Hamiltonian and the solution. 

In the next sections we present the explicit framework for designing solvable one-particle problems in continuous and discrete spaces and illustrate our strategy of inverse engineering with several nontrivial examples.

\section{Reconstruction of the real space potentials}\label{Sec-Real-space}

In this section we illustrate our strategy of generating the exact solutions for a simpler and more familiar case of a single quantum particle in the real space. Let us consider an electron in the three-dimensional space subjected to a time-dependent external electromagnetic field. It is convenient to use the temporal gauge in which the electric $\Ev$ and magnetic $\Bv$ fields are related to the vector potential $\Av(\rv,t)$ as follows \footnote{Throughout this article we work in the system of units in which $\hbar,c,e=1$.}
\begin{eqnarray}\label{B-E}
\Bv=\nabla \times \Av;~~~~ \Ev=-\partial _t \Av.
\end{eqnarray}
Given the vector potential $\Av(\rv,t)$ and the initial state  $\psi_0(\rv)=|\psi_0(\rv)|e^{\varphi_0(\rv)}$ the wave function $\psi(\rv,t)$ is obtained by solving the following \SE
\begin{equation} \label{SE-RT1} 
  \id \partial_t \psi (\rv,t)=\dfrac{1}{2m} \big( -i\nabla-     \Av(\rv,t)\big)^2	 \psi(\rv,t).
\end{equation}

Here we are interested in the vector potentials for which the \SE possesses an analytic solution. Therefore we follow our strategy and apply TDCDFT maps $\jv\mapsto\Av$ and $\jv\mapsto\psi$ from the time-dependent current $\jv(\rv,t)$ to the vector potential and the wave function. These maps can be easily found explicitly provided the initial state is nodeless $\psi_0(\rv)\neq 0 $, and the current $\jv(\rv,t)$ fulfills the condition
\begin{equation} \label{Con-R1}
  \int_{t_0} ^t \nabla \cdot \jv(\rv,t) \neq |\psi_0(\rv)|^2=n_0(\rv),
\end{equation}   
which ensures that the state remains nodeless in the course of evolution. Then $\Av(\rv,t)$ and the time-dependent wave function $\psi(\rv,t)$ are uniquely reconstructed from the given current \cite{TokatlyUni2011}
\begin{subequations} \label{Map-R1}
  \begin{eqnarray} 
    A[ \psi_0,\jv]&=& \nabla  \varphi(\rv,t)- m \vv(\rv,t),\label{A-RT1}\\
     \psi[ \psi_0,\jv]&=&\sqrt{n(\rv,t)}e^{\id \varphi(\rv,t)},\label{WF-RT1}\\    
       \varphi[ \psi_0,\jv] &=&  \varphi_0+ \int _{t_0} ^t  \d t' \bigg( \dfrac{\nabla^2 \sqrt{n(\rv,t')}}{2m \sqrt{n(\rv,t')}}-\frac{1}{2} m\vv^2(\rv,t) \bigg) \nonumber.\\\label{PH-RT1}
  \end{eqnarray}
\end{subequations}
where the density $n(\rv,t)$ and the velocity $ \vv(\rv,t)$ are defined as follows
\begin{subequations}
  \begin{eqnarray} 
    n(\rv,t)&=&-\int_{t_0} ^t \nabla \cdot \jv(\rv,t)+n_0(\rv),\label{D-R1}\\
    \vv(\rv,t)&=&\dfrac{\jv(\rv,t)}{n(\rv,t)}.\label{VE-R1}
  \end{eqnarray}   
\end{subequations}
One can straightforwardly check by a direct substitution that these formulas indeed give a solution to Eq.~\eqref{SE-RT1}. These equations allow us to reconstruct the driving potential for a prescribed evolution of the current.

Using Eqs. \eqref{Map-R1} with different time dependent currents we can construct infinitely many \SEs with time dependent vector potentials, which are all analytically solvable and the solutions are given by \Qs \eqref{WF-RT1} and \eqref{PH-RT1}.

Now we turn to a more specific situation when the system is driven by a longitudinal electric field $\Ev$ at zero magnetic field, $\Bv =0$. In the absence of the magnetic field $\nabla \times \Av=0$ and therefore \Q \eqref{A-RT1}  implies that the curl of the velocity also vanishes, $\nabla \times \vv=0 $. 
The velocity of a one-particle system driven by a potential field is must be potential.  In this case it is natural to use the Coulomb gauge ($\nabla \cdot \Av=0$) and express the electric field as a gradient of the scalar potential $V(\rv,t)$, that is $\Ev=-\nabla V$. By applying the standard gauge transformation to Eq.~\eqref{SE-RT1} we obtain the \SE of the following form
\begin{equation}\label{SE-RC}
  i \partial_t \psi (\rv,t)=\left(-\dfrac{\nabla^2}{2m}+V(\rv,t)\right) 	 \psi(\rv,t).
\end{equation}
where $\psi (\rv,t)$ is now the time-dependent wave function in the Coulomb gauge.

Applying the gauge transformation то \Qs \eqref{Map-R1} we find the mapping from the current $\jv(\rv,t)$ or equivalently the velocity and density $n(\rv,t)$ to the external potential $V(\rv,t)$ and  the wave function $\psi (\rv,t)$ \cite{MaiBurWoo2002}
\begin{subequations}\label{Map-R2}
  \begin{eqnarray}
    V(\rv,t)&=&\dfrac{\nabla ^2 \sqrt{n(\rv,t)} }{2m  \sqrt{n(\rv,t)}}- m \int_0 ^{\rv} \dot \vv(\rv',t) \cdot \d \rv' \nonumber\\
    &-&\dfrac{1}{2}m\vv^2 -\dot C(t) \label{V-RC1} \\ 
    \psi(\rv,t)&=&\sqrt{n(\rv,t)}e^{ \id\phi(\rv,t)}, \label{VF-RC1}\\
    \phi(\rv,t) &=& \int_0 ^\rv m\vv(\rv',t) \cdot \d \rv'+ C(t),\label{PH-RC1}
  \end{eqnarray}
\end{subequations}
where $C(t)$ is a time-dependent constant. Since by construction the velocity is irrotational the line integrals in Eqs.~\eqref{V-RC1} and \eqref{PH-RC1} do not depend on the integration path. Therefore we indicate only the initial and the final points of the path. The value $C(t_0)$ at the initial time is uniquely determined by the initial condition, while for $t>t_0$ the function $C(t)$ is arbitrary and can be chosen at convenience, for example to fix the value of the potential at infinity. The presence of a time-dependent constant in the density-potential mapping is in agreement with the Runge-Gross theorem \cite{RunGro1984}. The first term in Eq.~\eqref{V-RC1} is the Bohm potential that can be interpreted physically as an adiabatic potential for which the prescribed (nodeless) $n(\rv,t)$ is the instantaneous ground state density. The second and the third terms in Eq.~\eqref{V-RC1} are related to inertia forces. These terms compensate the  inertia forces exerted on a particle in a local non-inertial frame 
moving with the velocity $\vv(\rv,t)$. As a result in this co-moving frame the density stays stationary and equal to the initial density distribution. In the original frame the velocity-dependent contribution appears as a deformation of the adiabatic potential, which is aimed at supporting the prescribed density in the case of arbitrary fast evolution. 
 
There is an important difference of the present construction and the explicit current-vector potential mapping of Eqs.~\eqref{Map-R1}-\eqref{VE-R1}.
Equations \eqref{Map-R2} state that given  the density $n$ and the corresponding velocity $\vv$, the external  potential $V$ and the wave function $\psi$ can be found analytically. However the density and the velocity are not independent variables as they have to be consistent through the continuity equation 
\begin{equation}\label{Con-nv}
\dot n(\rv,t)=- \nabla\cdot\left[n(\rv,t) \vv(\rv,t)\right].
\end{equation}
The complication comes from the requirement of irrotational velocity, $\nabla \times \vv=0$ which implies the velocity field of the form $\vv=\nabla\Phi$. Because of this condition there is no a simple and universal analytic relation between the observables entering \eqref{Map-R2}. Such a relation can be found only for 1D systems, or if we assume a 1D inhomogeneity of the observables. In higher dimensions our ability of constructing solvable quantum problems is limited by the possibility to solve analytically a classical hydrodynamics problem of reconstructing the density from the velocity or vice versa for an irrotational flow. Below we present a simple example of such a reconstruction.

\subsection{Exact solution generated by a time-dependent scaling of observables}

Let the evolution starts from the ground state  $\psi_0$  of a potential  $V_0(\rv)$ with the  ground state density  $n_0(\rv)$ and the energy $E_0$. The simplest irrotational velocity field $\vv(\rv,t)$ for which Eq.~\eqref{Con-nv} can be solved analytically is a linear function of coordinates with time-dependent coefficients
\begin{equation}\label{velocity}
  \vv(\rv,t)=\dot \lambda(t)[ \rv- \rv_0(t)]+\dot \rv_0(t).
\end{equation}
This velocity corresponds to rigid motion of a fluid supplemented with a uniform expansion/compression relatively to the origin moving along the trajectory $\rv=\rv_0(t)$. The expansion/compression scaling factor is related to the parameter $\lambda(t)$ as, $\alpha(t)=e^{\lambda(t)}$. This interpretation is confirmed by solving the continuity equation \eqref{Con-nv} with the velocity of Eq.~\eqref{velocity}. The corresponding solution for the density takes the form 
\begin{equation} \label{D-R11}
  n(\rv,t)=\frac{1}{\alpha^3(t)}n_0\left(\frac{\rv-\rv _0(t)}{\alpha(t)}\right),
\end{equation}
which indeed corresponds to a rescaled density moving along the trajectory $\rv=\rv_0(t)$. The assumed initial conditions, $n(\rv,t_0)=n_0(\rv)$ and $\vv(\rv,t_0)=0$, are fulfilled if the time dependent parameters $\lambda$ and $\rv_0$ have zero values and zero time derivatives at the initial time, that is $\lambda(t_0)=0$, $\rv_0(t_0)=0$, $\dot\lambda(t_0)=0$ and $\dot \rv_0(t_0)=0$. 

Now we can insert the prescribed observables, Eqs.~\eqref{velocity} and \eqref{D-R11}, into Eq.~\eqref{Map-R2} to reconstruct the corresponding potential and the wave function.

To calculate the Bohm potential entering Eqs.~\eqref{V-RC1} we make use of the fact that $n_0 (\rv)$ is the ground state density of the potential $V_0(\rv)$ with the energy $E_0$. This implies that the shifted and rescaled density $n(\rv,t)$ of Eq.~\eqref{D-R11} corresponds to the instantaneous ground state of the shifted and rescaled potential $\alpha ^{-2} V_0 \left((\rv- \rv_0)/\alpha\right)$ with the ground state energy $\alpha^{-2} E_0$.
Therefore the Bohm potential can be represented as
\begin{equation}\label{Bohm}
\dfrac{\nabla^2 \sqrt{ n(\rv,t)} }{2m \sqrt{ n(\rv,t)}}=
\frac{1}{\alpha^2(t)} V_0 \left(\frac{\rv-\rv _0(t)}{\alpha(t)}\right)-\frac{1}{\alpha^2(t)}E_0.
\end{equation}
The final results for the potential and the wave function generated by the velocity \eqref{velocity} [or equivalently by the density \eqref{D-R11}] take the following form
\begin{subequations}\label{Map-R3}
  \begin{eqnarray}
   V(\rv,t) &=& \frac{1}{\alpha^2}V_0\left(\frac{\rv-\rv_0}{\alpha}\right)- m\ddot\rv_0\cdot\rv \nn \\
    &-& \frac{m}{2}\frac{\ddot\alpha}{\alpha}(\rv-\rv_0)^2, \label{V-RC2}    \\ 
     \psi(\rv,t) &=& \sqrt{\frac{1}{\alpha^3}n_0\left(\frac{\rv-\rv_0}{\alpha}\right)}e^{ \id\varphi(\rv,t)},\\
   \varphi(\rv,t)&=&\frac{m}{2} \frac{\dot \alpha}{\alpha}\left(\rv-\rv_0\right)^2+m\dot\rv_0 \cdot \rv \nn\\
&-&\int_0 ^t (\frac{1}{\alpha^2}E_0+\frac{m}{2}\dot \rv_0^2)dt' 
\label{PH-RC2} 
  \end{eqnarray}
\end{subequations}
Obviously, the first term in Eq.~\eqref{V-RC2} is the adiabatic potential. The other two terms describe two types of inertia forces -- the usual linear acceleration force (the second term) the inertial force related to a time-dependent deformation.  

In the special case of rigid motion, $\alpha=1$ or $\lambda=0$, only a linear acceleration inertial correction survives, so that the potential of Eq.~\eqref{V-RC1} simplifies as $V(\rv,t)=V_0(\rv-\rv_0)-m\ddot\rv_0\cdot\rv$. This potential rigidly transports a quantum system along a given trajectory without any reshaping of the initial density profile. It is worth noting that in this particular case our solution to the \SE is not limited to one particle and can be trivially generalized to a system of any number of interacting identical particles. Indeed, the solution generated by a spatially uniform velocity field $\vv(\rv,t)=\dot\rv_0(t)$ can be obtained by the transformation to a uniformly accelerated reference frame \cite{Vignale1995a}. Since the relative motion of particles is unaffected by this transformation the above potential will transport the center of mass while keeping unchanged the quantum state for the relative motion. It is absolutely obvious that if the initial state $\psi_0$ corresponds to 
that of the harmonic potential, our solution is identical to the harmonic potential theorem \cite{Dobson1994,Vignale1995a,Vignale1995b}.

One can also easily see that the analytic solution of the \SE for a harmonic oscillator with a time-dependent frequency $\omega(t)$ and a driving force ${\bf f}(t)$ \cite{PopPer1970}
\begin{equation} \label{SE-POPPER}
 \id \partial_t \psi (\rv,t)= \left(-\frac{\nabla^2}{2m}+\frac{1}{2}m\omega^2(t)\rv^2-{\bf f}(t)\cdot\rv\right)\psi (\rv,t),
\end{equation}
is a particular case of our Eq.~\eqref{Map-R3}. Assuming $V_0(\rv)=\frac{1}{2}m \omega_0^2\rv^2$ in Eq.~\eqref{V-RC2} we find that the reconstructed potential $V(\rv,t)$ coincides (up to irrelevant constant) with the potential in  Eq.~\eqref{SE-POPPER}, where
\begin{eqnarray}
&&\omega^2(t)= \frac{\omega_0^2}{\alpha^4} - \frac{\ddot\alpha}{\alpha},\\
&&{\bf f}(t)= m \left(\frac{\omega_0^2}{\alpha^4} - \frac{\ddot\alpha}{\alpha}\right)\rv_0 + m\ddot \rv_0.
\end{eqnarray}
From these two equations we observe that the center of mass position $\rv_0$ is the solution to the Newton equation for a driven harmonic oscillator
\begin{equation}
m\ddot \rv_0 + m \omega^2(t) \rv_0={\bf f}(t).
\end{equation}
Hence in this particular case the solution of \SE for the driven quantum oscillator is expressed in terms of the solution for the classical driven oscillator, which is the main observation made in Refs.~\onlinecite{PopPer1969,PopPer1970}.

\section{Inverse engineering of solvable quantum problems on a discrete space}\label{Sec-Discrete-space}
 
In this section we describe and illustrate our general reconstruction strategy for lattice systems. In this case we use the maps for a generalized lattice-TDCDFT \cite{TokatlyL2011} to inverse engineer analytically solvable one-particle problems (or problems isomorphic to one-particle dynamics on a lattice) \cite{TokatlyL2011}.

Our starting point is the \SE for the wave function $\psi_i(t_0)$ describing a particle on an $M$-site lattice with time-dependent complex hopping parameters $T_{ij}$,
\begin{equation} \label{SE-Chain1}
  \id \partial_t \psi_i(t)= - \sum _{j=1} ^M T_{ij}(t) ~\psi_j (t) ~~;~~T_{ii}=0,
\end{equation}
where indexes $i$ and $j$ take values on the lattice sites indicating the position in the discrete space, and $T_{ij}=T_{ji}^*$ to have a Hermitian Hamiltonian. 
In the \SE \eqref{SE-Chain1} we adopted a temporal gauge in which the scalar on-site potential and, possibly, a magnetic field enter via the phase of the hopping parameters $T_{ij}(t)$ \cite{TokatlyL2011}. For generality we also allow a time-dependent hopping rate $|T_{ij}(t)|$.

In the generalized lattice-TDCDFT of Ref.~\onlinecite{TokatlyL2011} the complex hopping $T_{ij}(t)$ plays a role of a driving potential. The corresponding observable in this approach can be called a "complex current" \cite{TokatlyL2011}
\begin{equation} \label{Q}
  Q_{ij}(t)=2 T_{ij}(t)\psi_i^*(t) \psi_j(t).
\end{equation} 
The real part of $Q_{ij}(t)$ is equal to physical current $J_{ij}$ on the lattice link connecting sites $i$ and $j$, while the real its real part $K_{ij}$ represents the kinetic energy on the link
\begin{equation}\label{Q1}
  Q_{ij}=K_{ij}+i J_{ij}.
\end{equation} 
The link current $J_{ij}(t)$ and the on-site density $n_i(t)=|\psi_i(t)|^2$ are connected by the lattice continuity equation
\begin{equation}
 \label{l-continuity}
\dot n_i(t) = -\sum_j J_{ij}(t).
\end{equation}
Since the link current and the link kinetic energy are, respectively, antisymmetric and symmetric with respect reversing the direction of the lattice link, $J_{ij}=-J_{ij}$ and $K_{ij}=K_{ij}$, the combined complex observable $Q_{ij}$ is a Hermitian matrix, $Q_{ij}=Q^*_{ij}$.

Given the complex current $Q_{ij}(t)$ and the initial state $\psi_i(t_0)=|\psi_i(t_0)|e^{\varphi_i(t_0)}\neq0$, the complex hopping $T_{ij}(t)$ and the wave function $\psi_i(t)=|\psi_i(t)|e^{\varphi_i(t)}$ can be expressed explicitly as functions of $Q_{ij}$ and $\psi_i(t_0)$
\begin{subequations}\label{Map-L1}
  \begin{eqnarray} 
    &&T_{ij}(t)=\dfrac{Q_{ij}(t)}{2 \psi_i^*(t)\psi_j(t)},\label{T}\\
    &&|\psi_i(t)|=\sqrt{|\psi_i(t_0)|^2 - \int_{t_0}^ t \sum_j J_{ij}(t') ~ \d t' },\label{|psi|}\\
    &&\varphi_i(t)=\varphi_i(t_0)+\int_{t_0}^t \dfrac{\sum_j K_{ij} (t') }{2 |\psi_i(t')|^2} \d t'.
  \end{eqnarray}
\end{subequations}
These formulas provide us with the analytic lattice-TDCDFT map from the observable to the conjugate driving potential and the corresponding solution of the time-dependent \SE. Using this map we can construct infinitely many analytically solvable problems generated by different time-dependent Hermitian observables $Q_{ij}(t)$.

Below we will give two examples which illustrate the possibility to analytically control quantum dynamics time in a discrete space. 

\subsection{Dynamics of one particle on a 1D chain}\label{Ssec-OP}

In this subsection we use our approach to manipulate the on-site density of a quantum particle on a finite tight-binding chain. Let us consider a particle on an atomic chain with a nearest neighbor hopping parameters $T_{i,i+1}$ of   of fixed amplitude $|T_{i,i+1}|=T_0$. The dynamics of the system is described by Eq.~\eqref{SE-Chain1}. Since for 1D systems only scalar (on-site) driving potentials are allowed, one can always gauge transformation the Hamiltonian to the form with real hopping parameters $T_{i,i+1}=T_0$ and the real on-site potential $v_i(t)$.\cite{TokatlyL2011} In the new gauge, which is the lattice analog of the Coulomb gauge, the time-dependent \SE reads:  
\begin{equation}\label{SE-Chain}
  \id \partial_t \psi_i = - T_0 (\psi_{i+1} + \psi_{i-1}) + v_i (t) \psi_i.
\end{equation}
In the following we assume for definiteness that the evolution starts from the ground state $\psi_0$ of the chain.

The equations for the observables \eqref{Q} and \eqref{Q1} for $Q_{ij}$ ,$J_{ij}$ and $K_{ij}$ remain the same except in the right hand side of  \Q \eqref{Q} the hopping $T_{ij}$ and the density matrix $\rho_{ij}$  need to be replaced by their counterparts in the Coulomb gauge.

The map \eqref{Map-L1} from the complex current $Q_{ij}$ to the hopping $T_{ij}$ and the wave function $\psi_i$ in the new gauge is transformed to an analytic map from $Q_{ij}$ to the on-site potential $v_i$ and the wave function $\psi_i$
\begin{subequations}
\begin{eqnarray}\label{Pot-chain}
       v_i=&&\sum_{j=i} ^M \Big( -\dfrac{K_{j,j+1}+K_{j-1,j}}{2n_j}+\dfrac{K_{j,j+1}+K_{j+1,j+2}}{2n_{j+1}}\nonumber  \\
 &&+\dfrac{\dot J _{j,j+1} K_{j,j+1}-J _{j,j+1} \dot K_{j,j+1}}{4T_0^2 n_i n_{j+1}} \Big), \label{V-chain1}\\
  \psi_i (t)&=& \sqrt{n_i (t)}e^{\id\varphi _i (t)},\label{WF-Chain}\\
  \varphi_i(t) &=& \varphi_i(t_0)+ \int_{t_0} ^t \big[\dfrac{K_{i,i+1}+K_{i-1,i}}{2n_i}-v_i\big] dt'.
\end{eqnarray} 
\end{subequations}
The important point is that in the considered physical situation with the fixed hopping amplitude the above formulas are not sufficient to reconstruct the potential from the given dynamics of observables. The fixed value of the hopping amplitude sets an upper bound on allowed values of link currents. As a result not all possible $Q$ become physically allowed, or $v$-representable in the TDDFT terminology. In fact, from the definition of Eq.~\eqref{Q} we find that the modulus of physically allowed $Q_{ij}(t)$ is bounded from above  
\begin{equation} \label{Q^2}
|Q_{ij}(t)|^2= 4T_0 ^2 n_i(t) n_j(t) < 4T_0 ^2.
\end{equation}

Formally the condition of the fixed hopping amplitude reduces the dimension of the space $\mathcal{N}$ of observables. In the present case this restriction can be taken into account by expressing $J_{ij}(t)$ and $K_{ij}(t)$ in terms of on-site density $n_i(t)$. Firstly, in 1D we can solve the continuity equation \eqref{l-continuity} to get the link current
\begin{equation}\label{J-Chain1}
 J_{i,i+1}=-\sum_{j=1}^i \dot n_j .
\end{equation}
Secondly, we express $K_{ij}(t)$ in terms of $n_i(t)$ using Eqs.~\eqref{Q^2} and \eqref{J-Chain1}
\begin{equation}
K_{i,i+1}=\pm \sqrt{4 T_0^2 n_i n_{i+1}- \Big(\sum_{j=1}^i \dot n_j\Big)^2}, \label{K-chain1}
\end{equation}
where the sign is determined by the sign of  $K_{ij}(t_0)$ at the initial time through the given initial state $\psi_0$.\cite{FarTok2012} Finally, by inserting Eqs.~\eqref{J-Chain1} and \eqref{K-chain1} into Eq.~\eqref{Pot-chain} we obtain the explicit analytic formulas for the reconstruction of the on-site lattice potential and the corresponding wave function from a given time-dependent density distribution. These formulas correspond to the maps of the lattice TDDFT \cite{LiUll2008,FarTok2012}. It is interesting to note that exact solution proposed recently in Ref.~\onlinecite{EdwDas2012} for a driven two-level system is, in fact, based on the above lattice TDDFT maps for a particular case of a two-site lattice.
 
\begin{figure}[ht] 
  \centering 
  \includegraphics[trim = 0mm 00mm 0mm 00mm, clip,width=0.5\textwidth]{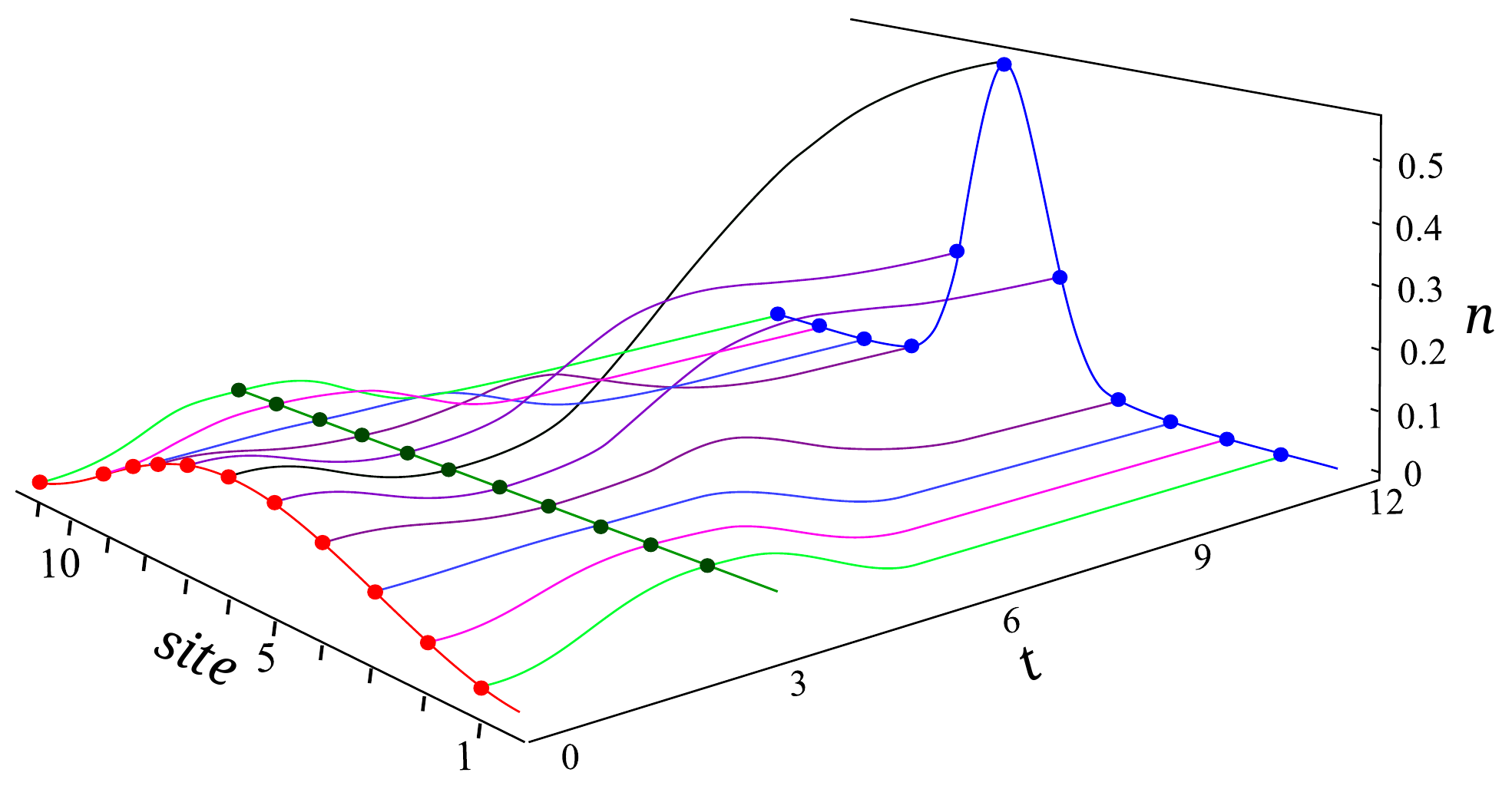}
  \caption{(Color online) The time evolution of the on-site densities determined by Eq.~\eqref{n-chain}. Each line represents the dynamic of the density in a particular site. The time duration for the first stage is $3$ units, $t_1=3$  and for the second stage is $9 $ units, $t_2=12$. Dots and their envelops indicate the density distribution in the initial $t=0$, intermediate ($t=3$), and the final ($t=12$) states.}
  \label{fig:Density1}
\end{figure}

Let us now demonstrate how this map works in practice by constructing a potential that a produces prescribed evolution of a density. Consider a particle on an atomic chain with 11 sites and a positive hopping constant $T_0=1$ and assume that the dynamics starts from the ground state of the chain with zero on-site potential,
\begin{equation}\label{GS-chain}
  \psi_i(0)=\psi_i ^{gs}= \frac{1}{\sqrt{6}} \sin \big(\frac{\pi i}{12}\big).
\end{equation} 
We will construct the driving potential which generates the following two-stage evolution: ({\it i}) On the first stage for $0<t<t_1$ the system evolves from the ground state of Eq.~\eqref{GS-chain} to a state with a homogeneous density distribution  $n_i=1/11$; ({\it ii}) On the second stage for $t_1<t<t_2$ the homogeneous density distribution shrinks to the center of the chain and by $t=t_2$ concentrates at site 6 with a Gaussian envelope, $n_i \propto e^{- (i-6)^2}$. The the required time evolution of density $n_i(t)$ for this two-stage process is the following, 
\begin{eqnarray}\label{n-chain}
  n_i(t) = \left\{ 
  \begin{array}{l l}
   \frac{1}{11}S(t/t_1 )+\big (1-S(t/t_1)\big)|\psi_i ^{gs}|^2 &  0 \leq t \leq t_1\\\\
 \mathcal{N}(t) \exp \left[-S\right(\frac{t-t_1}{t_2-t_1}\left) (i-6)^2 \right] & t_1<t \leq t_2
  \end{array}  \right.
\end{eqnarray}
where $\mathcal{N}(t)$ is the normalization factor
\begin{equation}
  \mathcal{N}(t)=\Big(\sum_{i=1} ^{11} e^{-S(\frac{t-t_1}{t_2-t_1}) (i-6)^2} \Big)^{-1}.
\end{equation}
Here $S(x)$ is a smooth step-like function which start from zero  at $x=0$ and reaches unity at $x=1$. For the reason that will be clear later we choose a function which has a zero first and second derivatives at $x=0,1$. Specifically here we use the following smooth step function which satisfies the above conditions conditions 
\begin{equation}\label{SF1}
  S(x)= x- \frac{1}{2 \pi} \sin (2\pi x).
\end{equation}
The time dependence of on-site densities $n_i(t)$ defined by Eq.~\eqref{n-chain} with $t_1=3$ and $t_2=12$ is shown on Fig.~\ref{fig:Density1}. Each line on the figure shows the prescribed evolution of the density on a particular site.  At $t=0$ the system is in the ground state \eqref{GS-chain}, then it goes gradually to the homogeneous distribution at $t=3$. Afterwards the density starts shrinking and finally at $t=12$ it reaches a bell shaped Gaussian centered at the middle site.

\begin{figure}[ht]
  \centering
  \includegraphics[trim = 0mm 00mm 0mm 0mm, clip,width=0.43\textwidth]{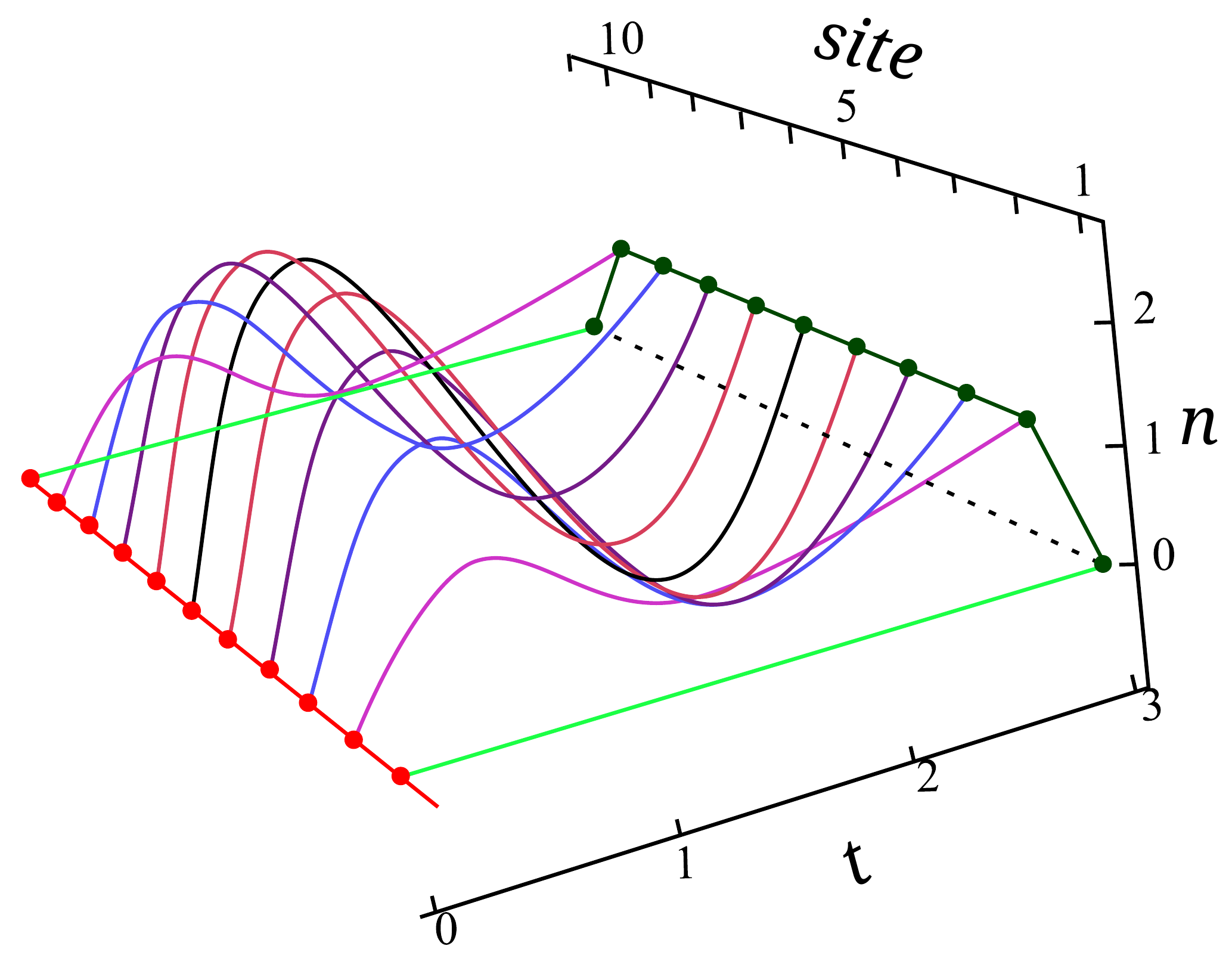}
  \caption{(Color online) On-site potentials \eqref{Pot-chain}, for the first stage of the evolution, $0\leq t \leq 3$, as a function of time. The on-site potentials for all sites are zero at the beginning. At $t=3$ all $v_i$ except those for the boundary sites reach $1$, while the potentials for the two ending points stay zero. Color online.}
   \label{fig:Potential1}
\end{figure}

\begin{figure} [ht]
  \centering
  \includegraphics[width=0.48\textwidth]{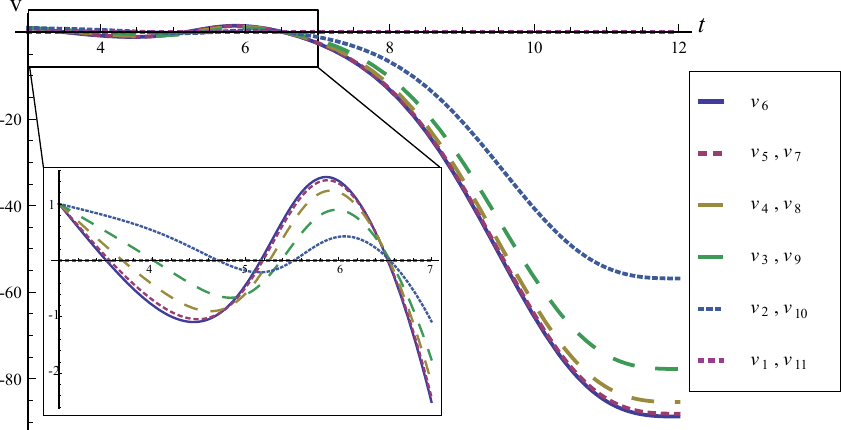}
  \caption{(Color online) On-site potentials \eqref{Pot-chain} as functions of time for the second stage of the evolution, $3\leq t \leq 12$. The evolution for $3<t<7$, is zoomed in the magnified box. }
  \label{fig:Potential2}
\end{figure}

The analytic representation for the corresponding driving potential can now be found immediately by inserting Eq.~\eqref{n-chain} into Eq.~\eqref{Pot-chain} where the link current $J_{ij}$ and the kinetic energy $K_{ij}$ are given by Eqs.~\eqref{J-Chain1} and \eqref{K-chain1}, respectively. and plugging in to the equation for the on-site potential $v_i$ \eqref{Pot-chain}. Since for the initial ground state $K_{ij}(0)$ is negative the minus sign must be chosen in Eq.~\eqref{K-chain1}.\cite{FarTok2012}  Fig.~\ref{fig:Potential1} shows the on-site potentials for the first stage of the dynamics, $0\leq t\leq 3 $. Each curve represents the time dependence of the potential for a particular site. Similarly, Fig.~\ref{fig:Potential2} shows the driving potential for the second stage of the time evolution. 


Our reconstructed potential has one interesting property. On the first stage of the evolution the potential shown on Fig.~\ref{fig:Potential2}  drives the system from its ground state at $t=0$ an takes it into a state with a homogeneous density at $t=3$ which is also the ground state of the system with the instantaneous potential $v_i(t=3)$. 
Similarly, for the second stage of the evolution the driving potential shown Fig.\ref{fig:Potential2} takes the system from ground state and brings it into a new ground state with a Gaussian envelope. At the first glance this behavior looks surprising as the dynamics is by far non-adiabatic. The explanation is, however, simple. The system at $t=t_1$ and $t=t_2$ is in its instantaneous ground state  because the first and second derivative of the step function $S$ \eqref{SF1} are zero at those times and therefore the current $J_{ij}$ and its first time derivative also vanish. By setting $J_{ij}=\dot J_{ij}=0$ in Eq.~\eqref{Pot-chain} we find the potential of the form
 \begin{equation}
  v_i=T_0 \sum _{j=i} ^M \left(\frac{\sqrt{n_j}+\sqrt{n_{j+2}}}{\sqrt{n_{j+1}}} -\frac{\sqrt{n_{j-1}}+\sqrt{n_{j+1}}}{\sqrt{n_j}}\right).
\end{equation} 
This is a lattice analog of the Bohm potential that corresponds to the ground state potential for a given instantaneous density. Therefore our construction can be used to make a fast transfer of a system between different ground states.

\subsection{Reconstruction of a driving magnetic field for a spin-1/2 system}\label{Ssec-SC}

As the last example we inverse engineer an analytically solvable \SE for a driven spin-1/2 system, or, equivalently, a generic two-level system. This formalism can be used, for example, to control the state evolution of a spin-1/2 using the time-dependent magnetic field. A similar problem has been addressed recently in Ref.~\onlinecite{Barnes2013}. Below we apply our general reconstruction strategy based on the lattice TDCDFT mapping.

Assume a spin in the initial state $|\psi_0\rangle$ subject to a time dependent magnetic filed $\Bv(t)$. The time dependent \SE for the state vector $|\psi(t)\rangle$ reads
\begin{equation} \label{SE-2S1}
  \id \partial_t |\psi(t) \rangle = - \Bv(t) \cdot  \hat \Sv ~|\psi(t) \rangle,
\end{equation}
where $\hat \Sv$ is the spin-1/2 operator.

By a gauge transformation one can always eliminate $z$-component of the magnetic filed and therefore reduce the problem to solving the \SE with the magnetic field in the $xy$-plane
\begin{equation}\label{SE-S2}
  \id \partial_t 
  \begin{pmatrix}
    \psi_\UP (t)\\
    \psi_\DN (t)\\
 \end{pmatrix}
=
-\begin{pmatrix}
   0 & \mathcal{B} \\
   \mathcal{B}^* & 0 \\
 \end{pmatrix}
 \begin{pmatrix}
   \psi_\UP (t) \\
   \psi_\DN (t)\\
 \end{pmatrix},
\end{equation}
where the "complex" magnetic field $\mathcal{B}$ is
\begin{equation} \label{B1}
  \mathcal{B}=( B_x-\id B_y)/2,
\end{equation}
and $\psi_\UP (t)=\langle\psi(t)|\UP\rangle$ and $\psi_\DN (t)=\langle\psi(t)|\DN\rangle$ are the projections of the state vector on the eigenstates of $\hat{S}_z$.

Equation \eqref{SE-S2} is identical to the \SE \eqref{SE-Chain1} for a two site lattice where the spin indexes $\UP$ and $\DN$ label the sites, and $\B(t)$ is the complex hopping parameter. Therefore we can directly apply the lattice-TDCDFT maps of Eqs.~\eqref{Map-L1} to reconstruct the driving magnetic field and the wave functions 
$\psi_{\UP,\DN}=|\psi_{\UP,\DN}|e^{\id\varphi_{\UP,\DN}}$ from given dynamics of the complex observable $Q(t)=K(t)+\id J(t)$ defined in Eq.~\eqref{Q}. In the present case the reconstruction formulas reduce to the form
\begin{subequations}\label{Map-S1}
  \begin{eqnarray} 
    &&|\psi_{\UP,\DN}(t)|=\sqrt{|\psi_{\UP,\DN}(0)|^2 \mp \int_{0}^ t \Im Q(t') ~ \d t' },\label{psi-S1}\\
    &&\varphi_{\UP,\DN}(t)=\varphi_{\UP,\DN}(0)+\int_{0}^t \dfrac{\Re Q (t') }{2 |\psi_{\UP,\DN}(t')|^2} \d t',\label{varphi_S1}\\
    &&\B(t)=\dfrac{Q(t)}{2 \psi_\UP ^*(t)\psi_\DN(t)}.\label{B-tilde}
  \end{eqnarray}
\end{subequations}
 
Equations \eqref{Map-S1} provide an analytic parametrization of the driving field and the wave function in terms of a given trajectory in the two-dimensional space $\mathcal{N}=\{K,J\}$ of observables. The point in the space $\mathcal{N}$ corresponds to a given kinetic energy $K$ and intersite current $J$ for a particle on the two-site lattice. This physical parametrization is universally applicable to lattice systems with any number of sites. In the particular two-site case one can propose an alternative parameterization the driving field, which has an intuitive interpretation in the physical context of the spin-1/2 system. Below we map the space $\mathcal{N}=\{K,J\}$ onto a Bloch sphere and rearrange Eqs.~\eqref{Map-S1} accordingly to relate the driving field to a given trajectory in the projective Hilbert space for spin-1/2.

As a first step we represent the state vector of spin-1/2 as follows
\begin{equation} \label{Projected-psi}
|\psi(t)\rangle= e^{\id\beta}\left[\cos(\theta/2)|\UP\rangle+e^{\id \phi} \sin(\theta/2) |\DN\rangle\right]
\end{equation}
where $\theta(t)$ and $\phi(t)$ are the spherical angles representing a point on the Bloch sphere, and $\beta(t)$ is an overall phase of the wave function. Next, to map the trajectory $Q(t)$ to a trajectory on the Bloch we use Eq.~\eqref{Map-S1} and express $Q$ in terms of the wave function amplitudes $|\psi_{\UP,\DN}|$ and the relative phase $\phi$
\begin{equation}
Q= 2\dot{\phi} \dfrac{|\psi_\UP|^2|\psi_\DN|^2}{|\psi_\UP|^2-|\psi_\DN|^2} -\id  \partial_t |\psi_\UP|^2 .\label{Q2}
\end{equation}
By substituting $|\psi_{\UP,\DN}|$ from Eq.~\eqref{Projected-psi} we relate the complex coordinate $Q$ in the space $\mathcal{N}$ to the spherical coordinates $(\theta,\phi)$ on the Bloch sphere
\begin{eqnarray} \label{Q-S2}
Q=\frac{1}{2}\left( \dot \phi \sin{\theta}\tan{\theta} + \id~ \dot \theta \sin \theta \right).
\end{eqnarray}  
This equation gives the required map between the trajectory in the original space of observables to the corresponding trajectory of spin-1/2 on the Bloch sphere. Finally, by inserting $Q$ of Eq.~\eqref{Q-S2} into Eq.~\eqref{B-tilde} for $\B$ \eqref{B-tilde} we get a new analytic representation for components of the magnetic field
\begin{subequations} \label{B2}
  \begin{eqnarray}
    B_x &=&  \dot \phi \tan{ \theta} \cos{ \phi}+ \dot\theta \sin{ \phi},\\
    B_y &=& -\dot{ \theta} \cos{ \phi} +\dot \phi \tan{ \theta} \sin{\phi}.
  \end{eqnarray}
\end{subequations}
The spherical coordinates $(\theta,\phi)$ determine the wave function Eq.~\eqref{Projected-psi} up to a common phase $\beta$. The phase $\beta=\varphi_\UP$ is calculated directly from Eq.~\eqref{varphi_S1} by substituting the expressions of $|\psi_\UP|$ and $Re Q$ in terms of $\theta$ and $\phi$,
\begin{equation}\label{Beta}
\beta (t)=\beta(0)+\frac{1}{2}\int_0 ^t\d t' \dot\phi\tan{\theta} \tan{\frac{\theta}{2}}.
\end{equation}
Equations \eqref{B2}, \eqref{Projected-psi}, and \eqref{Beta} solve the problem of reconstructing the driving field and the wave function from a given trajectory on the Bloch sphere.

Equations \eqref{B2} demonstrate one subtlety, which is very similar to the v-representability problem in TD(C)TDFT. Not all trajectories on the Bloch sphere are physically reproducible if the driving magnetic field is limited to the $(x,y)$-plane. For example, it is impossible to drive the system along the equator $(\pi/2,\phi(t))$ with a finite magnetic field because the right hand side in Eq.~\eqref{B2} diverges at $\theta=\pi/2$. Similarly, any trajectory which causes a divergence in right hand side of Eq.~\eqref{B2} is not $\mathcal{B}$-representable. All physically allowed trajectories when crossing the equator should approach it in a  way that $\dot\phi\tan\theta$ stays finite, which translates to the condition $\dot\phi\to 0$ when $\theta\to\pi/2 $. In other words, a physical trajectory, generated on the Bloch sphere by an in-plane magnetic field, can cross the line $\theta=\pi/2$ only if it is perpendicular to that line at the crossing point. This is absolutely clear physically because at any 
instant the magnetic filed generates rotation of the spin vector about the direction of ${\bf B}$. Therefore the initially coplanar to ${\bf B}$ spin is always driven out of the plane. Apparently when reconstructing the driving field from a trajectory on the Bloch sphere we should take it from a $\mathcal{B}$-representable set containing trajectories which either do not touch the equator or cross it perpendicularly.

Now we are ready to present an explicit example of the reconstruction.

\subsubsection*{Analytically controlled spin flip: design of a quantum NOT gate}

To illustrate our inverse engineering formulas we construct a control pulse which does the operation. Initially the system is in the ground state corresponding to some magnetic field ${\bf B}(0)$. During the pulse duration $\tau$ the magnetic field is changing and at the end of the pulse returns to its initial value ${\bf B}(\tau)={\bf B}(0)$ while the system is driven to the excited state in the field ${\bf B}(0)$. Therefore after the pulse the Hamiltonian returns to the initial form, but the direction of the spin is reversed.

For definiteness we assume the initial/final field in the $x$-direction, ${\bf B}(0)={\bf B}(\tau)=\hat{\xv}B_0$. Therefore the initial state is $|\leftarrow\rangle=1/\sqrt{2} (|\UP\rangle+|\DN \rangle)$. The target state which should be reached at the end of the pulse corresponds to another eigenstate of $\hat S_x$, that is 
$|\rightarrow\rangle=1/\sqrt{2}(|\UP\rangle-|\DN\rangle)$. On the Bloch sphere the initial and the final (target) states correspond to the points $(\pi/2, 0)$ and $(\pi/2, \pi)$, respectively.

The first step in constructing the required control pulse is find a $\mathcal{B}$-representable trajectory which at $t=0$ starts at the point $(\pi/2, 0)$ and arrives to the point $(\pi/2, \pi)$ at the time $t=\tau$. We note either boundary point belongs to the equatorial line. Therefore the trajectory should leave the initial point and arrive to the final point along the corresponding meridians. To automatically take care of the $\mathcal{B}$-representability we introduce a new independent variable $\gamma=\dot\phi\tan{\theta}$. The angle $\theta$ in the relevant range of $0<\theta<\pi$ is related to the new variable as follows
\begin{equation}
 \theta = 2\arctan\left[\sqrt{(\dot\phi/\gamma)^2+1}-\dot\phi/\gamma\right].
\end{equation}
Now the ``dangerous'' equatorial points correspond to the points of the trajectory with $\dot\phi/\gamma =0$.  

By re-expressing the magnetic field of Eqs.~\eqref{B2} and the common phase $\beta$ of Eq.~\eqref{Beta} in terms of $\gamma $ and $\phi$ we find
\begin{subequations} \label{B3}
  \begin{eqnarray}
    B_x &=&  \gamma \cos{ \phi} - \frac{\partial_t(\dot\phi/\gamma)}{(\dot\phi/\gamma)^2+1} \sin{\phi},\\
    B_y &=& \frac{\partial_t(\dot\phi/\gamma)}{(\dot\phi/\gamma)^2+1}\cos{\phi} + \gamma \sin{\phi}.
  \end{eqnarray}
\end{subequations}
and
\begin{equation}\label{Beta1}
\beta (t)=\beta(0)+\frac{1}{2}\int_0 ^t\d t' \left[\gamma\sqrt{(\dot\phi/\gamma)^2+1}-\dot\phi\right].
\end{equation}

Now we need to find two functions $\gamma(t)$ and $\phi(t)$ which will do the required job. The first obvious set of conditions for $\phi(t)$ is
\begin{equation}
 \phi(0)=0,\quad \phi(\tau)=\pi, \quad {\rm and} \quad \dot\phi(0)=\dot\phi(\tau)=0.
\end{equation}
It follows from Eq.~\eqref{B3} that the requirement $\Bv(0)=\Bv(\tau)=B_0\hat\xv$ will be fulfilled if
\begin{equation}
\gamma(0)=-\gamma(\tau)=B_0,
\end{equation}
and $\partial_t(\dot\phi/\gamma)=0$ at the boundary points, $t=0,\tau$. The latter condition is satisfied if at $t=0,\tau$ the second derivative of $\phi(t)$ vanishes
\begin{equation}
 \ddot \phi(0)=\ddot \phi(\tau)=0.
\end{equation}
In addition we have to make sure that the ratio $\dot\phi/\gamma$ is finite for all $0<t<\tau$.

As an example we suggest the following $\gamma(t)$ and $\phi(t)$ which fulfill all above conditions  
\begin{subequations}\label{gamma-phi}
  \begin{eqnarray}
    \gamma(t)&=&\frac{1}{2}B_{x0}(1 - 2t/\tau)\big[(1 - 2t/\tau)^2 + 3\big],\label{gamma}\\
    \varphi(t)&=& \frac{ \pi t}{\tau} - \frac{1}{4} \sin(\frac{4 \pi t}{\tau}). \label{phi}
  \end{eqnarray}
\end{subequations}
Here $\gamma(t)$ is a smooth monotonically decreasing function antisymmetric with respect to the point $t=\tau/2$. It goes from $B_0$ to $-B_0$ and crosses zero at the middle of the pulse. As an extra condition we required that $\dot\gamma(0)=\dot\gamma(\tau)=0$ which allows to smoothly continue the driving field beyond the interval $0<t<\tau$. The function $\phi(t)$ in Eq.~\eqref{phi} increases monotonically from $0$ to $\pi$ and has zero first and second derivatives at the boundary points $t=0$ and $t=\tau$, and at $t=\tau/2$. The derivative 
$\dot\phi(t)=\frac{2\pi t}{\tau}\sin^2(\frac{4\pi t}{\tau})$ is symmetric with respect to the middle point $t=\tau/2$. 

\begin{figure}
  \centering
  \includegraphics[trim = 0mm 20mm 0mm 0mm, clip,width=0.4\textwidth]{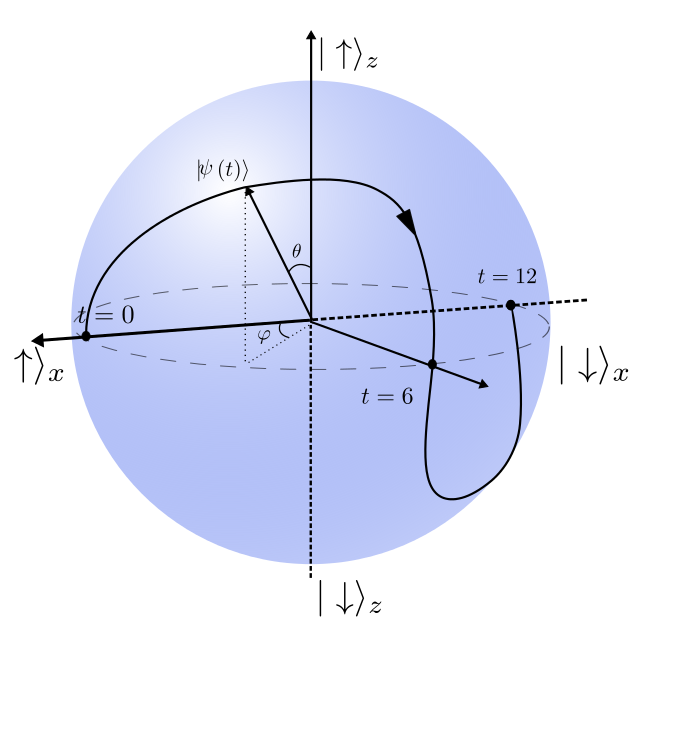}
  \caption{Trajectory of the spin state on the Bloch sphere generated by Eqs.~\eqref{gamma-phi} for $\tau=12$ and $B_{0}=1$. The trajectory leaves ground state $|\leftarrow\rangle$ and arrives to the excited state $|\rightarrow\rangle$ perpendicularly ($\dot \phi=0$) to the equatorial line.}
   \label{fig:Bloch}
\end{figure}

The corresponding trajectory on the Bloch-sphere is shown in Fig. \ref{fig:Bloch} for $B_0=1$ and $\tau=12$. The trajectory starts from the state $|\leftarrow\rangle$ on the equator, goes to the upper hemisphere, then at $t=\tau/2$ it crosses the equatorial line at the point $(\pi/2,\pi/2)$ and reaches the final state $|\rightarrow\rangle$ from the lower hemisphere. Because of the special symmetry of the generating functions the trajectory has a central symmetry with respect to the middle point.

\begin{figure}[h]
  \centering
  \includegraphics[trim = 0mm 00mm 0mm 0mm, clip,width=0.35\textwidth]{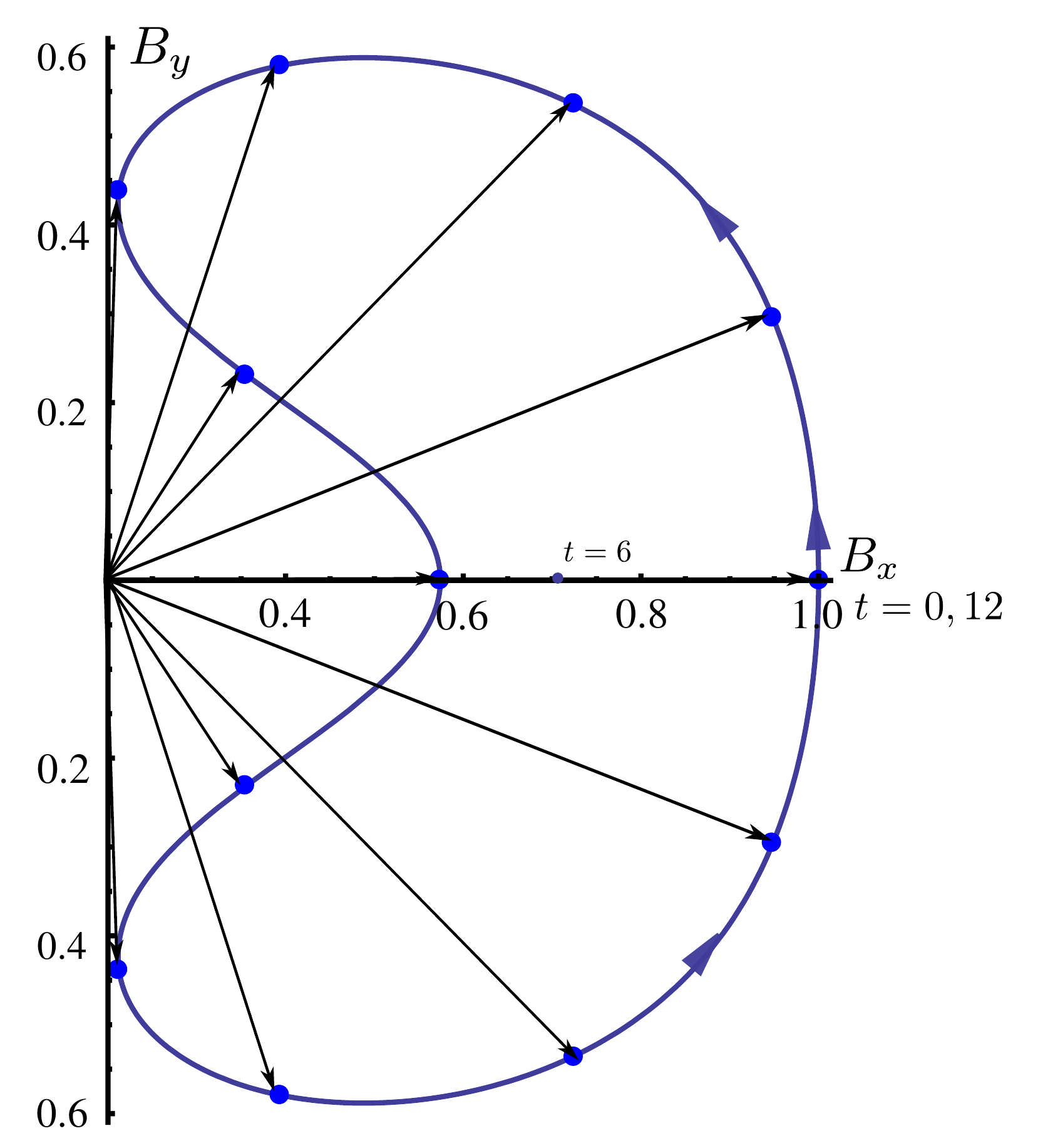}
  \caption{The trajectory traced out the magnetic field vector \eqref{B3} in the $xy$-plane. Starting from the initial value ${\bf B}(0)=\hat{\xv}$ the magnetic field follows the trajectory clockwise and comes back at $t=\tau=12$. Each arrow represents the magnetic field vector at integer times $t=0,1,\dots 12$.  }
  \label{fig:Magnetic-field}
\end{figure}

Substituting $\gamma(t)$ and $\phi(t)$ into Eqs. \eqref{B3} we find magnetic field generating this dynamics. In Fig. \ref{fig:Magnetic-field} we plot the path of the time-dependent magnetic field vector in the $xy$-plane. Each dot represents the magnetic field vector at integer times from 0 to 12. 


The corresponding wave function is given by Eq.~\eqref{Projected-psi} with the common phase $\beta(t)$ defined after Eq.~\eqref{Beta1}. For our particular example we easily find the overall phase at the end of the pulse, $\beta(\tau)=\beta(0)-\pi/2$.
\footnote{Because $\gamma(t)$ is antisymmetric and $\dot\phi(t)$ is symmetric with respect to $\tau/2$, the integral of the term with the square root vanishes. Therefore we are left with the result $\beta(\tau)=\beta(0) -[\phi(\tau)-\phi(0)]/2$.} Hence by the end of the pulse the system, initially in the ground state, is transferred to the excited state, and the wave function acquires a common phase shift of $\pi/2$.

One can check that if the pulse defined by Eqs.~\eqref{B3} and \eqref{gamma-phi} is applied to the initial excited state 
$|\psi(0)\rangle=|\rightarrow\rangle$, the system is driven to the ground state $|\leftarrow\rangle$. Therefore our pulse can be viewed as a realization of a quantum NOT gate which transforms (up to a common phase shift) $|\leftarrow\rangle$ into $|\rightarrow\rangle$, and 
$|\rightarrow\rangle$ into $|\leftarrow\rangle$.

\section{Conclusion}\label{Sec-Conclusion}

In conclusion, we proposed and worked out the strategy of using the TD(C)DFT observable-potential maps to inverse engineer analytically solvable time-dependent \SEs and thus to design analytic pulses for quantum control problems. We considered a number of situations in which the TD(C)DFT maps are known analytically. In all those cases the basic observables, such as the density or the current, can be used for a convenient and physically intuitive parametrization of the driving potential and the corresponding wave function. As the first pedagogical example we considered the control problem for the standard textbook \SE describing dynamics of a single quantum particle driven by a time-dependent electromagnetic field. In the most general case the driving vector potential and the solution to the \SE can be uniquely reconstructed from a given current. If the dynamics is restricted to be generated by a scalar potential, the latter can be reconstructed from the dynamics of the density or, equivalently, from the 
given (irrotational) velocity field. We have demonstrated how from the general reconstruction formulas one can recover the known analytic solutions of the time-dependent \SE for a driven harmonic oscillator with a time-dependent frequency \cite{PopPer1969,PopPer1970}.

In the second part of this work we applied our general strategy to a less obvious problem of quantum dynamics on lattices (discrete spaces). Here we used the analytic maps for the one-particle generalized lattice-TDCDFT \cite{TokatlyL2011} where the basic observables are the intersite current and the kinetic energy. As a first illustration of this setup we considered a manipulation of a one particle state on a finite tight-binding chain. We constructed an analytic driving field that generates a fast reshaping of the initial ground state density to the ground state density of another potential. In the second example we demonstrated the engineering of analytically solvable two level \SEs describing, in particular, dynamics of spin-1/2 system driven by a time-dependent magnetic field. Here we constructed a cyclic analytic control pulse which works as a quantum NOT gate, that is, it flips the direction of the spin for two basis stats. 

This work further develops and extends promising  applications of TDDFT to the quantum control.\citep{NieRugVan2013,NieRugVan2014} More importantly, it connects the ides of the density-potential mapping in TDDFT and TDCDFT to a wider range of coherent control \cite{Besonetal2012} and the state preparation problems in the quantum computing \cite{WuPipetal2011,BriCreetal2012,MalBasetal2013}. We hope this connection will be beneficial for either field.

\acknowledgments 
We acknowledge financial support  by the Spanish Grant FIS2013-46159-C3-1-P,  ``Grupos Consolidados UPV/EHU del Gobierno Vasco'' (Gant No. IT578-13) and  Air Force Office of Scientific Research (Grant No. FA2386-15-1-0006 AOARD).


\begin{thebibliography}{10}%
\makeatletter
\providecommand \@ifxundefined [1]{%
 \ifx #1\undefined \expandafter \@firstoftwo
 \else \expandafter \@secondoftwo
\fi
}%
\providecommand \@ifnum [1]{%
 \ifnum #1\expandafter \@firstoftwo
 \else \expandafter \@secondoftwo
\fi
}%
\providecommand \enquote [1]{``#1''}%
\providecommand \bibnamefont  [1]{#1}%
\providecommand \bibfnamefont [1]{#1}%
\providecommand \citenamefont [1]{#1}%
\providecommand\href[0]{\@sanitize\@href}%
\providecommand\@href[1]{\endgroup\@@startlink{#1}\endgroup\@@href}%
\providecommand\@@href[1]{#1\@@endlink}%
\providecommand \@sanitize [0]{\begingroup\catcode`\&12\catcode`\#12\relax}%
\@ifxundefined \pdfoutput {\@firstoftwo}{%
 \@ifnum{\z@=\pdfoutput}{\@firstoftwo}{\@secondoftwo}%
}{%
 \providecommand\@@startlink[1]{\leavevmode}%
 \providecommand\@@endlink[0]{}%
}{%
 \providecommand\@@startlink[1]{%
  \leavevmode
  \pdfstartlink
   attr{/Border[0 0 1 ]/H/I/C[0 1 1]}%
   user{/Subtype/Link/A<</Type/Action/S/URI/URI(#1)>>}%
  \relax
 }%
 \providecommand\@@endlink[0]{\pdfendlink}%
}%
\providecommand \url  [0]{\begingroup\@sanitize \@url }%
\providecommand \@url [1]{\endgroup\@href {#1}{\urlprefix}}%
\providecommand \urlprefix [0]{URL }%
\providecommand \Eprint[0]{\href }%
\@ifxundefined \urlstyle {%
  \providecommand \doi [1]{doi:\discretionary{}{}{}#1}%
}{%
  \providecommand \doi [0]{doi:\discretionary{}{}{}\begingroup
  \urlstyle{rm}\Url }%
}%
\providecommand \doibase [0]{http://dx.doi.org/}%
\providecommand \Doi[1]{\href{\doibase#1}}%
\providecommand \bibAnnote [3]{%
  \BibitemShut{#1}%
  \begin{quotation}\noindent
    \textsc{Key:}\ #2\\\textsc{Annotation:}\ #3%
  \end{quotation}%
}%
\providecommand \bibAnnoteFile [2]{%
  \IfFileExists{#2}{\bibAnnote {#1} {#2} {\input{#2}}}{}%
}%
\providecommand \typeout [0]{\immediate \write \m@ne }%
\providecommand \selectlanguage [0]{\@gobble}%
\providecommand \bibinfo [0]{\@secondoftwo}%
\providecommand \bibfield [0]{\@secondoftwo}%
\providecommand \translation [1]{[#1]}%
\providecommand \BibitemOpen[0]{}%
\providecommand \bibitemStop [0]{}%
\providecommand \bibitemNoStop [0]{.\EOS\space}%
\providecommand \EOS [0]{\spacefactor3000\relax}%
\providecommand \BibitemShut [1]{\csname bibitem#1\endcsname}%
\bibitem{LandauIII}%
  \BibitemOpen
  \bibfield{author}{%
  \bibinfo {author} {\bibfnamefont{L.~D.}\ \bibnamefont{Landau}}\ and\ \bibinfo
  {author} {\bibfnamefont{E.~M.}\ \bibnamefont{Lifshitz}},\ }%
  \emph{\bibinfo {title} {Quantum Mechanics: Non-Relativistic Theory}},\
  \bibinfo {edition} {4th}\ ed.,\ \bibinfo {series} {Course of Theoretical
  Physics}, Vol.~\bibinfo {volume} {3}\ (\bibinfo {publisher} {Pergamon},\
  \bibinfo {address} {Oxford},\ \bibinfo {year} {1977})%
  \bibAnnoteFile{NoStop}{LandauIII}%
\bibitem{Landau1932}%
  \BibitemOpen
  \bibfield{author}{%
  \bibinfo {author} {\bibfnamefont{L.}~\bibnamefont{Landau}},\ }%
  \bibfield{journal}{%
  \bibinfo {journal} {Phys. Z. Sowjetunion}\ }%
  \textbf{\bibinfo {volume} {2}},\ \bibinfo {pages} {46} (\bibinfo {year}
  {1932})%
  \bibAnnoteFile{NoStop}{Landau1932}%
\bibitem{Zener1932}%
  \BibitemOpen
  \bibfield{author}{%
  \bibinfo {author} {\bibfnamefont{C.}~\bibnamefont{Zener}},\ }%
  \bibfield{journal}{%
  \bibinfo {journal} {Proceedings of the Royal Society of London. Series A}\ }%
  \textbf{\bibinfo {volume} {137}},\ \bibinfo {pages} {696} (\bibinfo {year}
  {1932})%
  \bibAnnoteFile{NoStop}{Zener1932}%
\bibitem{Rabi1937}%
  \BibitemOpen
  \bibfield{author}{%
  \bibinfo {author} {\bibfnamefont{I.~I.}\ \bibnamefont{Rabi}},\ }%
  \bibfield{journal}{%
  \Doi{10.1103/PhysRev.51.652}{\bibinfo {journal} {Phys. Rev.}}\ }%
  \textbf{\bibinfo {volume} {51}},\ \bibinfo {pages} {652} (\bibinfo {year}
  {1937})%
  \bibAnnoteFile{NoStop}{Rabi1937}%
\bibitem{PopPer1969}%
  \BibitemOpen
  \bibfield{author}{%
  \bibinfo {author} {\bibfnamefont{V.}~\bibnamefont{Popov}}\ and\ \bibinfo
  {author} {\bibfnamefont{A.}~\bibnamefont{Perelomov}},\ }%
  \bibfield{journal}{%
  \bibinfo {journal} {Soviet Physics JETP}\ }%
  \textbf{\bibinfo {volume} {29}},\ \bibinfo {pages} {738} (\bibinfo {year}
  {1969})%
  \bibAnnoteFile{NoStop}{PopPer1969}%
\bibitem{PopPer1970}%
  \BibitemOpen
  \bibfield{author}{%
  \bibinfo {author} {\bibfnamefont{V.}~\bibnamefont{Popov}}\ and\ \bibinfo
  {author} {\bibfnamefont{A.}~\bibnamefont{Perelomov}},\ }%
  \bibfield{journal}{%
  \bibinfo {journal} {Soviet Physics JETP}\ }%
  \textbf{\bibinfo {volume} {30}},\ \bibinfo {pages} {910} (\bibinfo {year}
  {1970})%
  \bibAnnoteFile{NoStop}{PopPer1970}%
\bibitem{Dobson1994}%
  \BibitemOpen
  \bibfield{author}{%
  \bibinfo {author} {\bibfnamefont{J.~F.}\ \bibnamefont{Dobson}},\ }%
  \bibfield{journal}{%
  \bibinfo {journal} {Phys. Rev. Lett.}\ }%
  \textbf{\bibinfo {volume} {73}},\ \bibinfo {pages} {2244} (\bibinfo {year}
  {1994})%
  \bibAnnoteFile{NoStop}{Dobson1994}%
\bibitem{Vignale1995a}%
  \BibitemOpen
  \bibfield{author}{%
  \bibinfo {author} {\bibfnamefont{G.}~\bibnamefont{Vignale}},\ }%
  \bibfield{journal}{%
  \bibinfo {journal} {Phys. Rev. Lett.}\ }%
  \textbf{\bibinfo {volume} {74}},\ \bibinfo {pages} {3233} (\bibinfo {year}
  {1995})%
  \bibAnnoteFile{NoStop}{Vignale1995a}%
\bibitem{Vignale1995b}%
  \BibitemOpen
  \bibfield{author}{%
  \bibinfo {author} {\bibfnamefont{G.}~\bibnamefont{Vignale}},\ }%
  \bibfield{journal}{%
  \bibinfo {journal} {Phys. Lett. A}\ }%
  \textbf{\bibinfo {volume} {209}},\ \bibinfo {pages} {206} (\bibinfo {year}
  {1995})%
  \bibAnnoteFile{NoStop}{Vignale1995b}%
\bibitem{EcoSopetal2006}%
  \BibitemOpen
  \bibfield{author}{%
  \bibinfo {author} {\bibfnamefont{S.~E.}\ \bibnamefont{Economou}}, \bibinfo
  {author} {\bibfnamefont{L.~J.}\ \bibnamefont{Sham}}, \bibinfo {author}
  {\bibfnamefont{Y.}~\bibnamefont{Wu}},\ and\ \bibinfo {author}
  {\bibfnamefont{D.~G.}\ \bibnamefont{Steel}},\ }%
  \bibfield{journal}{%
  \Doi{10.1103/PhysRevB.74.205415}{\bibinfo {journal} {Phys. Rev. B}}\ }%
  \textbf{\bibinfo {volume} {74}},\ \bibinfo {pages} {205415} (\bibinfo {year}
  {2006})%
  \bibAnnoteFile{NoStop}{EcoSopetal2006}%
\bibitem{GreiEcoetal2009}%
  \BibitemOpen
  \bibfield{author}{%
  \bibinfo {author} {\bibfnamefont{A.}~\bibnamefont{Greilich}}, \bibinfo
  {author} {\bibfnamefont{S.~E.}\ \bibnamefont{Economou}}, \bibinfo {author}
  {\bibfnamefont{S.}~\bibnamefont{Spatzek}}, \bibinfo {author}
  {\bibfnamefont{D.~R.}\ \bibnamefont{Yakovlev}}, \bibinfo {author}
  {\bibfnamefont{D.}~\bibnamefont{Reuter}}, \bibinfo {author}
  {\bibfnamefont{A.~D.}\ \bibnamefont{Wieck}}, \bibinfo {author}
  {\bibfnamefont{T.~L.}\ \bibnamefont{Reinecke}},\ and\ \bibinfo {author}
  {\bibfnamefont{M.}~\bibnamefont{Bayer}},\ }%
  \bibfield{journal}{%
  \bibinfo {journal} {Nat Phys}\ }%
  \textbf{\bibinfo {volume} {5}},\ \bibinfo {pages} {262} (\bibinfo {year}
  {2009}),\ ISSN \bibinfo {issn} {1745-2473}%
  \bibAnnoteFile{NoStop}{GreiEcoetal2009}%
\bibitem{PoeKenetal2011}%
  \BibitemOpen
  \bibfield{author}{%
  \bibinfo {author} {\bibfnamefont{E.}~\bibnamefont{Poem}}, \bibinfo {author}
  {\bibfnamefont{O.}~\bibnamefont{Kenneth}}, \bibinfo {author}
  {\bibfnamefont{Y.}~\bibnamefont{Kodriano}}, \bibinfo {author}
  {\bibfnamefont{Y.}~\bibnamefont{Benny}}, \bibinfo {author}
  {\bibfnamefont{S.}~\bibnamefont{Khatsevich}}, \bibinfo {author}
  {\bibfnamefont{J.~E.}\ \bibnamefont{Avron}},\ and\ \bibinfo {author}
  {\bibfnamefont{D.}~\bibnamefont{Gershoni}},\ }%
  \bibfield{journal}{%
  \Doi{10.1103/PhysRevLett.107.087401}{\bibinfo {journal} {Phys. Rev. Lett.}}\
  }%
  \textbf{\bibinfo {volume} {107}},\ \bibinfo {pages} {087401} (\bibinfo {year}
  {2011})%
  \bibAnnoteFile{NoStop}{PoeKenetal2011}%
\bibitem{WuPipetal2011}%
  \BibitemOpen
  \bibfield{author}{%
  \bibinfo {author} {\bibfnamefont{Y.}~\bibnamefont{Wu}}, \bibinfo {author}
  {\bibfnamefont{I.~M.}\ \bibnamefont{Piper}}, \bibinfo {author}
  {\bibfnamefont{M.}~\bibnamefont{Ediger}}, \bibinfo {author}
  {\bibfnamefont{P.}~\bibnamefont{Brereton}}, \bibinfo {author}
  {\bibfnamefont{E.~R.}\ \bibnamefont{Schmidgall}}, \bibinfo {author}
  {\bibfnamefont{P.~R.}\ \bibnamefont{Eastham}}, \bibinfo {author}
  {\bibfnamefont{M.}~\bibnamefont{Hugues}}, \bibinfo {author}
  {\bibfnamefont{M.}~\bibnamefont{Hopkinson}},\ and\ \bibinfo {author}
  {\bibfnamefont{R.~T.}\ \bibnamefont{Phillips}},\ }%
  \bibfield{journal}{%
  \Doi{10.1103/PhysRevLett.106.067401}{\bibinfo {journal} {Phys. Rev. Lett.}}\
  }%
  \textbf{\bibinfo {volume} {106}},\ \bibinfo {pages} {067401} (\bibinfo
  {month} {Feb}\ \bibinfo {year} {2011})%
  \bibAnnoteFile{NoStop}{WuPipetal2011}%
\bibitem{BriCreetal2012}%
  \BibitemOpen
  \bibfield{author}{%
  \bibinfo {author} {\bibfnamefont{R.~T.}\ \bibnamefont{Brierley}}, \bibinfo
  {author} {\bibfnamefont{C.}~\bibnamefont{Creatore}}, \bibinfo {author}
  {\bibfnamefont{P.~B.}\ \bibnamefont{Littlewood}},\ and\ \bibinfo {author}
  {\bibfnamefont{P.~R.}\ \bibnamefont{Eastham}},\ }%
  \bibfield{journal}{%
  \Doi{10.1103/PhysRevLett.109.043002}{\bibinfo {journal} {Phys. Rev. Lett.}}\
  }%
  \textbf{\bibinfo {volume} {109}},\ \bibinfo {pages} {043002} (\bibinfo {year}
  {2012})%
  \bibAnnoteFile{NoStop}{BriCreetal2012}%
\bibitem{MalBasetal2013}%
  \BibitemOpen
  \bibfield{author}{%
  \bibinfo {author} {\bibfnamefont{N.}~\bibnamefont{Malossi}}, \bibinfo
  {author} {\bibfnamefont{M.~G.}\ \bibnamefont{Bason}}, \bibinfo {author}
  {\bibfnamefont{M.}~\bibnamefont{Viteau}}, \bibinfo {author}
  {\bibfnamefont{E.}~\bibnamefont{Arimondo}}, \bibinfo {author}
  {\bibfnamefont{R.}~\bibnamefont{Mannella}}, \bibinfo {author}
  {\bibfnamefont{O.}~\bibnamefont{Morsch}},\ and\ \bibinfo {author}
  {\bibfnamefont{D.}~\bibnamefont{Ciampini}},\ }%
  \bibfield{journal}{%
  \Doi{10.1103/PhysRevA.87.012116}{\bibinfo {journal} {Phys. Rev. A}}\ }%
  \textbf{\bibinfo {volume} {87}},\ \bibinfo {pages} {012116} (\bibinfo {month}
  {Jan}\ \bibinfo {year} {2013})%
  \bibAnnoteFile{NoStop}{MalBasetal2013}%
\bibitem{MotGametal2009}%
  \BibitemOpen
  \bibfield{author}{%
  \bibinfo {author} {\bibfnamefont{F.}~\bibnamefont{Motzoi}}, \bibinfo {author}
  {\bibfnamefont{J.~M.}\ \bibnamefont{Gambetta}}, \bibinfo {author}
  {\bibfnamefont{P.}~\bibnamefont{Rebentrost}},\ and\ \bibinfo {author}
  {\bibfnamefont{F.~K.}\ \bibnamefont{Wilhelm}},\ }%
  \bibfield{journal}{%
  \Doi{10.1103/PhysRevLett.103.110501}{\bibinfo {journal} {Phys. Rev. Lett.}}\
  }%
  \textbf{\bibinfo {volume} {103}},\ \bibinfo {pages} {110501} (\bibinfo
  {month} {Sep}\ \bibinfo {year} {2009})%
  \bibAnnoteFile{NoStop}{MotGametal2009}%
\bibitem{Economou2012}%
  \BibitemOpen
  \bibfield{author}{%
  \bibinfo {author} {\bibfnamefont{S.~E.}\ \bibnamefont{Economou}},\ }%
  \bibfield{journal}{%
  \Doi{10.1103/PhysRevB.85.241401}{\bibinfo {journal} {Phys. Rev. B}}\ }%
  \textbf{\bibinfo {volume} {85}},\ \bibinfo {pages} {241401} (\bibinfo {month}
  {Jun}\ \bibinfo {year} {2012})%
  \bibAnnoteFile{NoStop}{Economou2012}%
\bibitem{ChoDicetal2010}%
  \BibitemOpen
  \bibfield{author}{%
  \bibinfo {author} {\bibfnamefont{J.~M.}\ \bibnamefont{Chow}}, \bibinfo
  {author} {\bibfnamefont{L.}~\bibnamefont{DiCarlo}}, \bibinfo {author}
  {\bibfnamefont{J.~M.}\ \bibnamefont{Gambetta}}, \bibinfo {author}
  {\bibfnamefont{F.}~\bibnamefont{Motzoi}}, \bibinfo {author}
  {\bibfnamefont{L.}~\bibnamefont{Frunzio}}, \bibinfo {author}
  {\bibfnamefont{S.~M.}\ \bibnamefont{Girvin}},\ and\ \bibinfo {author}
  {\bibfnamefont{R.~J.}\ \bibnamefont{Schoelkopf}},\ }%
  \bibfield{journal}{%
  \Doi{10.1103/PhysRevA.82.040305}{\bibinfo {journal} {Phys. Rev. A}}\ }%
  \textbf{\bibinfo {volume} {82}},\ \bibinfo {pages} {040305} (\bibinfo {month}
  {Oct}\ \bibinfo {year} {2010})%
  \bibAnnoteFile{NoStop}{ChoDicetal2010}%
\bibitem{GamMotetal2011}%
  \BibitemOpen
  \bibfield{author}{%
  \bibinfo {author} {\bibfnamefont{J.~M.}\ \bibnamefont{Gambetta}}, \bibinfo
  {author} {\bibfnamefont{F.}~\bibnamefont{Motzoi}}, \bibinfo {author}
  {\bibfnamefont{S.~T.}\ \bibnamefont{Merkel}},\ and\ \bibinfo {author}
  {\bibfnamefont{F.~K.}\ \bibnamefont{Wilhelm}},\ }%
  \bibfield{journal}{%
  \Doi{10.1103/PhysRevA.83.012308}{\bibinfo {journal} {Phys. Rev. A}}\ }%
  \textbf{\bibinfo {volume} {83}},\ \bibinfo {pages} {012308} (\bibinfo {year}
  {2011})%
  \bibAnnoteFile{NoStop}{GamMotetal2011}%
\bibitem{BamBer1981}%
  \BibitemOpen
  \bibfield{author}{%
  \bibinfo {author} {\bibfnamefont{A.}~\bibnamefont{Bambini}}\ and\ \bibinfo
  {author} {\bibfnamefont{P.~R.}\ \bibnamefont{Berman}},\ }%
  \bibfield{journal}{%
  \bibinfo {journal} {Phys. Rev. A}\ }%
  \textbf{\bibinfo {volume} {23}},\ \bibinfo {pages} {2496} (\bibinfo {year}
  {1981})%
  \bibAnnoteFile{NoStop}{BamBer1981}%
\bibitem{BamArtetal1984}%
  \BibitemOpen
  \bibfield{author}{%
  \bibinfo {author} {\bibfnamefont{A.}~\bibnamefont{Bambini}}\ and\ \bibinfo
  {author} {\bibfnamefont{M.}~\bibnamefont{Lindberg}},\ }%
  \bibfield{journal}{%
  \bibinfo {journal} {Phys. Rev. A}\ }%
  \textbf{\bibinfo {volume} {30}},\ \bibinfo {pages} {794} (\bibinfo {year}
  {1984})%
  \bibAnnoteFile{NoStop}{BamArtetal1984}%
\bibitem{Hioe1984}%
  \BibitemOpen
  \bibfield{author}{%
  \bibinfo {author} {\bibfnamefont{F.~T.}\ \bibnamefont{Hioe}},\ }%
  \bibfield{journal}{%
  \bibinfo {journal} {Phys. Rev. A}\ }%
  \textbf{\bibinfo {volume} {30}},\ \bibinfo {pages} {2100} (\bibinfo {year}
  {1984})%
  \bibAnnoteFile{NoStop}{Hioe1984}%
\bibitem{Zakarzewski1985}%
  \BibitemOpen
  \bibfield{author}{%
  \bibinfo {author} {\bibfnamefont{J.}~\bibnamefont{Zakrzewski}},\ }%
  \bibfield{journal}{%
  \bibinfo {journal} {Phys. Rev. A}\ }%
  \textbf{\bibinfo {volume} {32}},\ \bibinfo {pages} {3748} (\bibinfo {year}
  {1985})%
  \bibAnnoteFile{NoStop}{Zakarzewski1985}%
\bibitem{SilJosHou1985}%
  \BibitemOpen
  \bibfield{author}{%
  \bibinfo {author} {\bibfnamefont{M.~S.}\ \bibnamefont{Silver}}, \bibinfo
  {author} {\bibfnamefont{R.~I.}\ \bibnamefont{Joseph}},\ and\ \bibinfo
  {author} {\bibfnamefont{D.~I.}\ \bibnamefont{Hoult}},\ }%
  \bibfield{journal}{%
  \Doi{10.1103/PhysRevA.31.2753}{\bibinfo {journal} {Phys. Rev. A}}\ }%
  \textbf{\bibinfo {volume} {31}},\ \bibinfo {pages} {2753} (\bibinfo {year}
  {1985})%
  \bibAnnoteFile{NoStop}{SilJosHou1985}%
\bibitem{Robinson1985}%
  \BibitemOpen
  \bibfield{author}{%
  \bibinfo {author} {\bibfnamefont{E.~J.}\ \bibnamefont{Robinson}},\ }%
  \bibfield{journal}{%
  \Doi{10.1103/PhysRevA.31.3986}{\bibinfo {journal} {Phys. Rev. A}}\ }%
  \textbf{\bibinfo {volume} {31}},\ \bibinfo {pages} {3986} (\bibinfo {year}
  {1985})%
  \bibAnnoteFile{NoStop}{Robinson1985}%
\bibitem{Ishkhanyan2000}%
  \BibitemOpen
  \bibfield{author}{%
  \bibinfo {author} {\bibfnamefont{A.~M.}\ \bibnamefont{Ishkhanyan}},\ }%
  \bibfield{journal}{%
  \bibinfo {journal} {Journal of Physics A: Mathematical and General}\ }%
  \textbf{\bibinfo {volume} {33}},\ \bibinfo {pages} {5539} (\bibinfo {year}
  {2000})%
  \bibAnnoteFile{NoStop}{Ishkhanyan2000}%
\bibitem{KyoVit2005}%
  \BibitemOpen
  \bibfield{author}{%
  \bibinfo {author} {\bibfnamefont{E.~S.}\ \bibnamefont{Kyoseva}}\ and\
  \bibinfo {author} {\bibfnamefont{N.~V.}\ \bibnamefont{Vitanov}},\ }%
  \bibfield{journal}{%
  \Doi{10.1103/PhysRevA.71.054102}{\bibinfo {journal} {Phys. Rev. A}}\ }%
  \textbf{\bibinfo {volume} {71}},\ \bibinfo {pages} {054102} (\bibinfo {year}
  {2005})%
  \bibAnnoteFile{NoStop}{KyoVit2005}%
\bibitem{Vitanov20007}%
  \BibitemOpen
  \bibfield{author}{%
  \bibinfo {author} {\bibfnamefont{N.~V.}\ \bibnamefont{Vitanov}},\ }%
  \bibfield{journal}{%
  \bibinfo {journal} {New Journal of Physics}\ }%
  \textbf{\bibinfo {volume} {9}},\ \bibinfo {pages} {58} (\bibinfo {year}
  {2007})%
  \bibAnnoteFile{NoStop}{Vitanov20007}%
\bibitem{GanDzeGal2010}%
  \BibitemOpen
  \bibfield{author}{%
  \bibinfo {author} {\bibfnamefont{A.}~\bibnamefont{Gangopadhyay}}, \bibinfo
  {author} {\bibfnamefont{M.}~\bibnamefont{Dzero}},\ and\ \bibinfo {author}
  {\bibfnamefont{V.}~\bibnamefont{Galitski}},\ }%
  \bibfield{journal}{%
  \Doi{10.1103/PhysRevB.82.024303}{\bibinfo {journal} {Phys. Rev. B}}\ }%
  \textbf{\bibinfo {volume} {82}},\ \bibinfo {pages} {024303} (\bibinfo {year}
  {2010})%
  \bibAnnoteFile{NoStop}{GanDzeGal2010}%
\bibitem{EdwDas2012}%
  \BibitemOpen
  \bibfield{author}{%
  \bibinfo {author} {\bibfnamefont{E.}~\bibnamefont{Barnes}}\ and\ \bibinfo
  {author} {\bibfnamefont{S.}~\bibnamefont{Das~Sarma}},\ }%
  \bibfield{journal}{%
  \Doi{10.1103/PhysRevLett.109.060401}{\bibinfo {journal} {Phys. Rev. Lett.}}\
  }%
  \textbf{\bibinfo {volume} {109}},\ \bibinfo {pages} {060401} (\bibinfo {year}
  {2012})%
  \bibAnnoteFile{NoStop}{EdwDas2012}%
\bibitem{Barnes2013}%
  \BibitemOpen
  \bibfield{author}{%
  \bibinfo {author} {\bibfnamefont{E.}~\bibnamefont{Barnes}},\ }%
  \bibfield{journal}{%
  \Doi{10.1103/PhysRevA.88.013818}{\bibinfo {journal} {Phys. Rev. A}}\ }%
  \textbf{\bibinfo {volume} {88}},\ \bibinfo {pages} {013818} (\bibinfo {month}
  {Jul}\ \bibinfo {year} {2013})%
  \bibAnnoteFile{NoStop}{Barnes2013}%
\bibitem{BanCheMugShe}%
  \BibitemOpen
  \bibfield{author}{%
  \bibinfo {author} {\bibfnamefont{X.~M. J. G. S. E.~Y.}\ \bibnamefont{Ban},
  \bibfnamefont{Yue;~Chen}}\ }%
  \textbf{\bibinfo {volume} {arXiv:1309.1916}} (\bibinfo {year} {2013})%
  \bibAnnoteFile{NoStop}{BanCheMugShe}%
\bibitem{Ishkhanyan32000}%
  \BibitemOpen
  \bibfield{author}{%
  \bibinfo {author} {\bibfnamefont{A.~M.}\ \bibnamefont{Ishkhanyan}},\ }%
  \bibfield{journal}{%
  \bibinfo {journal} {Journal of Physics A: Mathematical and General}\ }%
  \textbf{\bibinfo {volume} {33}},\ \bibinfo {pages} {5041} (\bibinfo {year}
  {2000})%
  \bibAnnoteFile{NoStop}{Ishkhanyan32000}%
\bibitem{RunGro1984}%
  \BibitemOpen
  \bibfield{author}{%
  \bibinfo {author} {\bibfnamefont{E.}~\bibnamefont{Runge}}\ and\ \bibinfo
  {author} {\bibfnamefont{E.~K.~U.}\ \bibnamefont{Gross}},\ }%
  \bibfield{journal}{%
  \Doi{10.1103/PhysRevLett.52.997}{\bibinfo {journal} {Phys. Rev. Lett.}}\ }%
  \textbf{\bibinfo {volume} {52}},\ \bibinfo {pages} {997} (\bibinfo {year}
  {1984})%
  \bibAnnoteFile{NoStop}{RunGro1984}%
\bibitem{TDDFT-2012}%
  \BibitemOpen
  \emph{\bibinfo {title} {Fundamentals of Time-Dependent Density Functional
  Theory}},\ edited by\ \bibinfo {editor} {\bibfnamefont{F.~M. N. E.~G.}\
  \bibnamefont{Miguel A.L.~Marques}, \bibfnamefont{Neepa T.~Maitra}}\ and\
  \bibinfo {editor} {\bibfnamefont{A.}~\bibnamefont{Rubio}}\ (\bibinfo
  {publisher} {springer},\ \bibinfo {year} {2012})%
  \bibAnnoteFile{NoStop}{TDDFT-2012}%
\bibitem{Ullrich-book}%
  \BibitemOpen
  \bibfield{author}{%
  \bibinfo {author} {\bibfnamefont{C.~A.}\ \bibnamefont{Ullrich}},\ }%
  \emph{\bibinfo {title} {Time-Dependent Density-Functional Theory: Concepts
  and Applications}}\ (\bibinfo {publisher} {Oxford University Press},\
  \bibinfo {year} {2012})%
  \bibAnnoteFile{NoStop}{Ullrich-book}%
\bibitem{DreizlerGross1990}%
  \BibitemOpen
  \bibfield{author}{%
  \bibinfo {author} {\bibfnamefont{R.~M.}\ \bibnamefont{Dreizler}}\ and\
  \bibinfo {author} {\bibfnamefont{E.~K.~U.}\ \bibnamefont{Gross}},\ }%
  \emph{\bibinfo {title} {Density-Functional Theory}}\ (\bibinfo {publisher}
  {Springer},\ \bibinfo {address} {Berlin},\ \bibinfo {year} {1990})%
  \bibAnnoteFile{NoStop}{DreizlerGross1990}%
\bibitem{NieRugVan2013}%
  \BibitemOpen
  \bibfield{author}{%
  \bibinfo {author} {\bibfnamefont{S.~E.~B.}\ \bibnamefont{Nielsen}}, \bibinfo
  {author} {\bibfnamefont{M.}~\bibnamefont{Ruggenthaler}},\ and\ \bibinfo
  {author} {\bibfnamefont{R.}~\bibnamefont{van Leeuwen}},\ }%
  \bibfield{journal}{%
  \bibinfo {journal} {EPL (Europhysics Letters)}\ }%
  \textbf{\bibinfo {volume} {101}},\ \bibinfo {pages} {33001} (\bibinfo {year}
  {2013})%
  \bibAnnoteFile{NoStop}{NieRugVan2013}%
\bibitem{NieRugVan2014}%
  \BibitemOpen
  \bibfield{author}{%
  \bibinfo {author} {\bibfnamefont{S.}~\bibnamefont{Nielsen}}, \bibinfo
  {author} {\bibfnamefont{M.}~\bibnamefont{Ruggenthaler}},\ and\ \bibinfo
  {author} {\bibfnamefont{R.}~\bibnamefont{van Leeuwen}},\ }%
  \bibfield{journal}{%
  \bibinfo {journal} {arXiv preprint arXiv:1412.3794}}%
   (\bibinfo {year} {2014})%
  \bibAnnoteFile{NoStop}{NieRugVan2014}%
\bibitem{MaiBurWoo2002}%
  \BibitemOpen
  \bibfield{author}{%
  \bibinfo {author} {\bibfnamefont{N.~T.}\ \bibnamefont{Maitra}}, \bibinfo
  {author} {\bibfnamefont{K.}~\bibnamefont{Burke}},\ and\ \bibinfo {author}
  {\bibfnamefont{C.}~\bibnamefont{Woodward}},\ }%
  \bibfield{journal}{%
  \bibinfo {journal} {Phys. Rev. Lett.}\ }%
  \textbf{\bibinfo {volume} {89}},\ \bibinfo {pages} {023002} (\bibinfo {year}
  {2002})%
  \bibAnnoteFile{NoStop}{MaiBurWoo2002}%
\bibitem{LiUll2008}%
  \BibitemOpen
  \bibfield{author}{%
  \bibinfo {author} {\bibfnamefont{Y.}~\bibnamefont{Li}}\ and\ \bibinfo
  {author} {\bibfnamefont{C.~A.}\ \bibnamefont{Ullrich}},\ }%
  \bibfield{journal}{%
  \Doi{10.1063/1.2955733}{\bibinfo {journal} {J. Chem. Phys.}}\ }%
  \textbf{\bibinfo {volume} {129}},\ \bibinfo {eid} {044105} (\bibinfo {year}
  {2008})%
  \bibAnnoteFile{NoStop}{LiUll2008}%
\bibitem{TokatlyUni2011}%
  \BibitemOpen
  \bibfield{author}{%
  \bibinfo {author} {\bibfnamefont{I.~V.}\ \bibnamefont{Tokatly}},\ }%
  \bibfield{journal}{%
  \Doi{http://dx.doi.org/10.1016/j.chemphys.2011.04.005}{\bibinfo {journal}
  {Chemical Physics}}\ }%
  \textbf{\bibinfo {volume} {391}},\ \bibinfo {pages} {78 } (\bibinfo {year}
  {2011}),\ ISSN \bibinfo {issn} {0301-0104}%
  \bibAnnoteFile{NoStop}{TokatlyUni2011}%
\bibitem{TokatlyL2011}%
  \BibitemOpen
  \bibfield{author}{%
  \bibinfo {author} {\bibfnamefont{I.~V.}\ \bibnamefont{Tokatly}},\ }%
  \bibfield{journal}{%
  \Doi{10.1103/PhysRevB.83.035127}{\bibinfo {journal} {Phys. Rev. B}}\ }%
  \textbf{\bibinfo {volume} {83}},\ \bibinfo {pages} {035127} (\bibinfo {month}
  {Jan}\ \bibinfo {year} {2011})%
  \bibAnnoteFile{NoStop}{TokatlyL2011}%
\bibitem{FarTok2012}%
  \BibitemOpen
  \bibfield{author}{%
  \bibinfo {author} {\bibfnamefont{M.}~\bibnamefont{Farzanehpour}}\ and\
  \bibinfo {author} {\bibfnamefont{I.~V.}\ \bibnamefont{Tokatly}},\ }%
  \bibfield{journal}{%
  \Doi{10.1103/PhysRevB.86.125130}{\bibinfo {journal} {Phys. Rev. B}}\ }%
  \textbf{\bibinfo {volume} {86}},\ \bibinfo {pages} {125130} (\bibinfo {month}
  {Sep}\ \bibinfo {year} {2012})%
  \bibAnnoteFile{NoStop}{FarTok2012}%
\bibitem{Note1}%
  \BibitemOpen
  \bibinfo {note} {Throughout this article we work in the system of units in
  which $\hbar ,c,e=1$.}%
  \bibAnnoteFile{Stop}{Note1}%
\bibitem{Note2}%
  \BibitemOpen
  \bibinfo {note} {Because $\gamma (t)$ is antisymmetric and $\protect
  \mathaccentV {dot}05F\phi (t)$ is symmetric with respect to $\tau /2$, the
  integral of the term with the square root vanishes. Therefore we are left
  with the result $\beta (\tau )=\beta (0) -[\phi (\tau )-\phi (0)]/2$.}%
  \bibAnnoteFile{Stop}{Note2}%
\bibitem{Besonetal2012}%
  \BibitemOpen
  \bibfield{author}{%
  \bibinfo {author} {\bibfnamefont{M.~G.}\ \bibnamefont{Bason}}, \bibinfo
  {author} {\bibfnamefont{M.}~\bibnamefont{Viteau}}, \bibinfo {author}
  {\bibfnamefont{N.}~\bibnamefont{Malossi}}, \bibinfo {author}
  {\bibfnamefont{P.}~\bibnamefont{Huillery}}, \bibinfo {author}
  {\bibfnamefont{E.}~\bibnamefont{Arimondo}}, \bibinfo {author}
  {\bibfnamefont{D.}~\bibnamefont{Ciampini}}, \bibinfo {author}
  {\bibfnamefont{R.}~\bibnamefont{Fazio}}, \bibinfo {author}
  {\bibfnamefont{V.}~\bibnamefont{Giovannetti}}, \bibinfo {author}
  {\bibfnamefont{R.}~\bibnamefont{Mannella}},\ and\ \bibinfo {author}
  {\bibfnamefont{O.}~\bibnamefont{Morsch}},\ }%
  \bibfield{journal}{%
  \bibinfo {journal} {Nature Physics}\ }%
  \textbf{\bibinfo {volume} {8}},\ \bibinfo {pages} {147} (\bibinfo {year}
  {2012})%
  \bibAnnoteFile{NoStop}{Besonetal2012}%
\end{thebibliography}
%
\end{document}